\documentclass[final,5p,times, fleqn]{elsarticle}

\usepackage{amssymb}
\usepackage{graphicx}
\usepackage{color}
\usepackage{listings}
\usepackage{amsmath}
\usepackage{url}

\newcommand{\fourier}[1]{\mathcal{F} \biggl[#1 \biggl]}
\newcommand{\ifourier}[1]{\mathcal{F}^{-1} \biggl[#1 \biggl]}

\newcommand{\fft}[1]{FFT \biggl[#1 \biggl]} 
\newcommand{\ifft}[1]{FFT^{-1} \biggl[#1 \biggl]} 

\begin{document}

\begin{frontmatter}



\title{Computational wave optics library for C++: CWO++ library}


\author[chibau]{Tomoyoshi Shimobaba\corref{cor1}}
\ead{shimobaba@faculty.chiba-u.jp}
\author[chibau]{Jian Tong Weng}
\author[chibau]{Takahiro Sakurai}
\author[chibau]{Naohisa Okada}
\author[chibau]{Takashi Nishitsuji}
\author[shohoku]{Naoki Takada}
\author[kisarazu]{Atsushi Shiraki}
\author[chibau]{Nobuyuki Masuda}
\author[chibau]{Tomoyoshi Ito}

\cortext[cor1]{Corresponding author}
\address[chibau]{Department of Artificial Systems, Graduate School of Engineering, Chiba University, 1-33 Yayoi-cho, Inage-ku, Chiba, Chiba 263-8522, Japan}
\address[shohoku]{Department of Informatics and Media Technology, Shohoku College, 428 Nurumizu, Atsugi, Kanagawa, 243--8501 Japan}
\address[kisarazu]{Department of Imformation and Computer Engineering, Kisarazu National College of Technology, Kiyomi-dai Higashi 2-11-1, Kisarazu, Chiba, 292-0041 Japan}

\begin{abstract}
Diffraction calculations, such as the angular spectrum method, and Fresnel diffractions, are used for calculating scalar light propagation.
The calculations are used in wide-ranging optics fields: for example, computer generated holograms (CGHs), digital holography, diffractive optical elements, microscopy, image encryption and decryption, three-dimensional analysis for optical devices and so on.
However, increasing demands made by large-scale diffraction calculations have rendered the computational power of recent computers insufficient.
We have already developed a numerical library for diffraction calculations using a graphic processing unit (GPU), which was named the GWO library.
However, this GWO library is not user-friendly, since it is based on C language and was also run only on a GPU.
In this paper, we develop a new C++ class library for diffraction and CGH calculations, which is referred as to a CWO++ library, running on a CPU and GPU.
We also describe the structure, performance, and usage examples of the CWO++ library.
\end{abstract}


\end{frontmatter}



\section{Introduction}

Scalar light propagation is calculated using several diffraction calculations.
These calculations, e.g. the angular spectrum method and Fresnel diffractions, are used in wide-ranging optics fields \cite{goodman, okan}, ultrasonic \cite{ultrasonic}, X-ray \cite{xray} and so on.
In optics, its applications include Computer Generated Holograms (CGH), digital holography, phase retrieval, image encryption and decryption, steganography, three-dimensional (3D) analysis for optical devices, Diffractive Optical Elements (DOE), and so on.

In CGH and digital holography, diffraction calculations are vital.
CGH are generated by calculating a diffraction calculation from a 3D object to a hologram plane on a computer.
If we apply CGH to a 3D display technique \cite{slinger, bentonbook}, the technique becomes attractive because a wavefront reconstructed from CGH is almost equivalent to an object light.
However, the computational time required for the diffraction calculation involved in CGH hampers the realization of a practical 3D display using CGH.
Many methods have thus been proposed for accelerating the computational time \cite{nlut, sakamoto, image_hol, ExtreamCGH, FraunhoferCGH}.

Digital holography is a well-known method for electronically recording existing 3D object information on a hologram, which is captured by a Charge Coupled Device (CCD) and Complementary Metal Oxide Semiconductor (CMOS) cameras \cite{schnars1, schnars2}.
To reconstruct 3D object information from a hologram in a computer, we need to calculate diffraction calculations.
Applications of digital holography include Digital Holographic Microscopy (DHM) \cite{kim, DHM2}, Digital Holographic Particle Tracking Velocimetry (DHPIV) \cite{masuda1, satake} and so forth.

Phase retrieval algorithm retrieves phase information of an object light from intensity patterns captured by CCD camera.
In optics, Gerchberg-Saxton (GS) algorithm \cite{GS} and modified algorithms, e.g. Fresnel ping-pong \cite{PingPong} and Yang-Gu \cite{YG} algorithms, are widely used for phase retrieval.
The algorithms are based on iterative optimization: namely, they gradually retrieve phase information by calculating diffraction calculations between certain intensity patterns (normally more than two) while subject to amplitude and phase constraints.
The applications of the algorithms include, for example, wavefront reconstruction \cite{wave_reconst1,wave_reconst2,wave_reconst3}, holographic projection \cite{proj1, proj2, proj3, proj4, shimo_proj1, shimo_proj2} and so on.

Diffraction based encryption and decryption \cite{encryption1, encryption2, encryption3}, and steganography \cite{steganography} were proposed .
Diffraction-based techniques have an interesting feature, and handle not only 2D but also 3D images.

In 3D analysis for optical devices, such as optical Micro Electro Mechanical Systems (MEMS), DOE and so on, we can obtain 3D light distribution to stack multiple two-dimensional (2D) diffraction calculations along depth-direction \cite{3d}.
For example, several research works have analyzed the optical characteristics of a Digital Micro-mirror Device (DMD), which is one of the MEMS devices, using Fresnel diffraction and the angular spectrum method \cite{mems, mems2}.

As mentioned, diffraction calculations are practically used in wide-ranging optics fields. The former can also accelerate computational time using the Fast Fourier Transform (FFT) algorithm; however, if we wish to realize real-time 3D reconstruction from holograms in digital holography, generate a large-area CGH, and obtain 3D light field from an optical device, recent computers lack sufficient computational power.

Using hardware is an effective means to further boost computational speed for CGH and diffraction calculations. 
In fact, we showed dramatically increased computational power to design and build special-purpose computers for CGH targeting a 3D display and named HOlographic ReconstructioN (HORN), in order to overcome the computational cost of CGH. 
The HORN computers designed by pipeline architecture can calculate light intensities on CGH at high speed \cite{horn1, horn2, horn3, horn4, horn5, horn6}.
The HORN computers were implemented on a Field Programmable Gate Array (FPGA) board, except HORN-1 and -2.
To date, we have constructed six HORN computers, which have been able to attain several thousand times the computational speed of recent computers.
Researchers also developed a special-purpose computer, FFT-HORN, in order to accelerate Fresnel diffraction in DHPIV \cite{masuda1, masuda2}.
The FFT-HORN was able to reconstruct $256 \times 256$ images from holograms, which were captured by a DHPIV optical system, at a rate of 30 frames per second.
The FPGA-based approaches for both CGH and diffraction calculations showed excellent computational speed, but are subject to the following restrictions: the high cost of developing the FPGA board, long development term, long compilation of the hardware description language and mapping to FPGA times, and technical know-how required for the FPGA technology.

Conversely, recent GPUs allow us to use as a highly parallel processor, because the GPUs have many simple processors, which can process 32- or 64-bit floating-point addition, multiplication and multiply-add instructions. 
The approach of accelerating numerical simulations using a GPU chip is referred to as General-Purpose computation on GPU (GPGPU) or GPU computing.
The merits of GPGPU include its high computational power, the low cost of the GPU board, short compilation time, and the short development term. 

We have already developed a numerical library for diffraction calculations using a GPU, which was named the GWO library \cite{gwo}.
The purpose of the GWO library is to facilitate access to GPU calculation power for optics engineers and researchers lacking GPGPU.
The GWO library has already been distributed via the Internet and used to report some papers.
For example, we reported on a real-time DHM system \cite{realtimeDHM}, diffraction calculations in a computer-aided design tool for developing a holographic display \cite{cad}, a fast CGH calculation \cite{wrp1,wrp2} and a DHM observable in multi-view and multi-resolution \cite{multiDHM}.
Moreover, researchers studied Airy beams generation and their propagation feature in the simulation using the GWO library \cite{airy1, airy2}.
However, the GWO library is not user-friendly because it is based on C language, e.g. the library user must manage the CPU and GPU memory allocation personally and so on.
In addition, the library is run on only a GPU, namely the diffraction and CGH calculations are not calculated on a CPU.

In this paper, we develop a C++ class library for computational wave optics involved in diffraction and CGH calculations, which is referred to as a CWO++ library, running on CPU and GPU.
The CWO++ library, unlike the GWO library, is developed using C++ and its 
structure, performance, and usage examples are described.

In Section 2, we describe diffraction calculations, the structure of the CWO++ library, the class ``CWO" for diffraction calculations, and the subclass ``cwoPLS" for Point Light Source (PLS)-based diffraction and CGH calculations on a CPU.
In Section 3, we describe the ``GWO" and ``gwoPLS" classes for diffraction and CGH calculations on a GPU.
In Section 4, we describe the implementation of the ``CWO" and ``GWO" classes.
In Section 5, we describe field types which are held in the classes.
In section 6, we show the performance of diffraction and CGH calculations on a CPU and GPU.
In Section 7, we show the applications of the CWO++ library to holography.
In Section 8, we conclude this work.

\section{Detail of the CWO++ library}

\begin{table*}
\caption{Classes of the CWO++ library. They are distributed as open-source codes.}
\begin{center}
\begin{tabular}{|c|c|c|c|}
\hline
Class & Role & Parent class & Related source files \\ \hline
cwo & \shortstack{Diffraction calculation \\ on CPU} & None & \shortstack{cwo.h\\ cwo.pp} \\ \hline
gwo & \shortstack{Diffraction calculation \\ on GPU} & cwo & \shortstack{gwo.h\\ gwo.cpp} \\ \hline
cwoPLS & \shortstack{PLS-based diffraction \\ and CGH calculations \\ on CPU} & cwo & \shortstack{cwoPLS.h\\ cwoPLS.cpp} \\ \hline
gwoPLS & \shortstack{PLS-based diffraction \\ and CGH calculations \\ on GPU} & gwo & \shortstack{gwoPLS.h\\ gwoPLS.cpp} \\ \hline
cwoComplex & Complex number & None & cwo.h \\ \hline
cwoObjPoint & PLS & None & cwo.h \\ \hline
cwoVect & Vector operations & None & cwo.h \\ \hline
\end{tabular}
\end{center}
\label{tbl:cwo}
\end{table*}

The CWO++ library mainly consists of two C++ classes: CWO and GWO.
The CWO class calculates diffraction calculations on a CPU, has auxiliary functions and allows us to calculate the following diffractions:
\begin{enumerate}
\item Fresnel diffraction(Convolution form)
\item Fresnel diffraction(Fourier form)
\item Shifted Fresnel diffraction
\item Angular spectrum method
\item Shifted angular spectrum method
\item 3D diffraction calculation 
\end{enumerate}
The first to fifth diffractions are primary diffraction calculations, on which the sixth diffraction calculation is based. The above diffraction calculations can also be calculated on a GPU using the GWO class. 

Table \ref{tbl:cwo} shows the structure of the CWO++ library.
The class ``CWO" is the top class of the CWO++ library, while the other classes, which are ``GWO", ``cwoPLS" and ``gwoPLS", are inherited from ``CWO".
The ``cwoPLS" and ``gwoPLS" classes are for PLS-based diffraction and CGH calculations on a CPU and GPU, respectively.
 
``cwoComplex" and ``cwoObjPoint" are data structures and their auxiliary functions for complex number and object points for PLS, respectively and are distributed as open-source codes.

\subsection{Diffraction calculation}

\begin{figure}[htb]
\centerline{
\includegraphics[width=8cm]{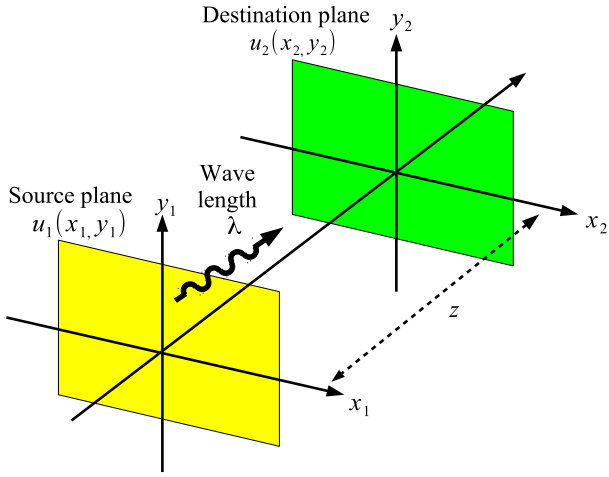}}
\caption{Diffraction calculation by monochromatic wave, whose wavelength is $\lambda$, between a source plane (aperture function) $u_1(x_1,y_1)$ and a destination plane $u_2(x_2, y_2)$.}
\label{fig:diff}
\end{figure}

Figure \ref{fig:diff} shows a diffraction calculation by monochromatic wave, whose wavelength is $\lambda$, between a source plane (aperture function) $u_1(x_1,y_1)$ and a destination plane $u_2(x_2, y_2)$. 

The CWO++ library allows us to calculate FFT-based diffraction calculations.
In addition, diffraction calculations are categorized into convolution and Fourier forms.
The former category includes Fresnel diffraction (convolution form), Shifted Fresnel diffraction, Angular spectrum method and Shifted angular spectrum method.
The latter category includes Fresnel diffraction (Fourier form).
In the following subsections, we describe these diffractions.

\subsubsection{Fresnel diffraction (convolution form)}

The convolution form of Fresnel diffraction is expressed by:
\begin{equation}
\begin{aligned}
u_2(x_2,y_2) & =  \frac{\exp(i \frac{2 \pi}{\lambda} z)}{i \lambda z}  
\int \!\!\int_{-\infty}^{+\infty} u_1(x_1, y_1) \\
& \exp(i \frac{\pi}{\lambda z} ((x_2-x_1)^2+(y_2-y_1)^2) ) dx_1 dy_1\end{aligned}
\label{eqn:fre_conv}
\end{equation}

The above equation is the convolution form, and can be expressed relative to the following equation according to convolution theorem:
\begin{eqnarray}
u_2(x_2,y_2) = \frac{\exp(i \frac{2 \pi}{\lambda} z)}{i \lambda z}  \ifourier{ \fourier{u(x_1,y_1)} \fourier{h_{F}(x_1,y_1)}}
\label{eqn:fresnel_conv}
\end{eqnarray}
where, the operators $\mathcal{F}[\cdot]$ and $\mathcal{F}^{-1}[\cdot]$ indicate Fourier and inverse Fourier transforms, respectively, 
$h_{F}(x,y)$ is the impulse response function (also known as the point spread function) of Eq. (\ref{eqn:fre_conv}) as follows,

\begin{eqnarray}
h_{F}(x,y) = \exp(i \frac{\pi}{\lambda z} (x^2+y^2) )
\label{eqn:fre_conv_prop}
\end{eqnarray}

In the numerical implementation, we need to discretize each spatial variable and use FFT instead of Fourier transforms as follows:
The discretizing space variables are $(x_1,y_1)=(m_1 \Delta x_1,\\
 n_1 \Delta y_1)$, where $\Delta x_1$ and $\Delta y_1$ are the sampling pitches and $m_1, n_1$ are integer indices on the source plane.
The discretizing space variables are $(x_2,y_2)=(m_2 \Delta x_2, n_2 \Delta y_2)$, where $\Delta x_2$ and $\Delta y_2$ are the sampling pitches and $m_2, n_2$ are integer indices on the destination plane.
The ranges of integer indices are as follows:

\begin{eqnarray}
&&-\frac{N_x}{2} \le m_1, m_2 < -\frac{N_x}{2}-1, \nonumber \\
&&-\frac{N_y}{2} \le m_1, m_2 < -\frac{N_y}{2}-1
\end{eqnarray}
where, $N_x$ and $N_y$ are the numbers of horizontal and vertical pixels on the source and destination planes, respectively.

The discretized Fresnel diffraction of the convolution form is as follows:
\begin{equation}
\begin{aligned}
u_2(m_2,n_2) &=  \frac{\exp(i \frac{2 \pi}{\lambda} z)}{i \lambda z} 
\ifft{ \fft{u(m_1,m_1)} \\
& \fft{h_{F}(m_1,m_1)}}
\end{aligned}
\label{eqn:fre_conv_fft}
\end{equation}

\begin{eqnarray}
h_{F}(m,n) = \exp(i \frac{\pi}{\lambda z} ((m \Delta x_1)^2+(n \Delta y_1)^2) )
\label{eqn:fre_conv_prop_d}
\end{eqnarray}

Note that the sampling pitches on the destination planes are the same as those on the source plane after the diffraction, namely $\Delta x_2=\Delta x_1$ and $\Delta y_2=\Delta y_1$.

\subsubsection{The Fresnel diffraction (Fourier form)}

We can obtain the Fourier form of the Fresnel diffraction to expand the quadratic term in Eq. (\ref{eqn:fre_conv}).
The form is expressed by:
\begin{equation}
\begin{aligned}
 & u(x_2,y_2)  =
\frac{\exp(i \frac{2 \pi}{\lambda} z)}{i \lambda z} 
\exp(i \frac{\pi}{\lambda z} (x_2^2+y_2^2)) \\
& \int \!\!\int_{-\infty}^{+\infty} 
u'(x_1,y_1) \exp(-i \frac{2 \pi}{\lambda z} (x_2 x_1 + y_2 x_1))
dx_1 dy_1,
\end{aligned}
\label{eqn:fre_fourier1}
\end{equation}
where $u'(x_1,y_1)$ is defined as,
\begin{eqnarray}
u'(x_1, y_1)= u(x_1, y_1) \exp(i \pi \frac{(x_1^2+y_1^2)}{\lambda z}).
\label{eqn:fre_fourier2}
\end{eqnarray}

This form is rewritten by the Fourier transform.
The discretized Fourier form of Fresnel diffraction is expressed by:
\begin{equation}
\begin{aligned}
u(m_2,n_2) = &
\frac{\exp(i \frac{2 \pi}{\lambda} z)}{i \lambda z}  
\exp(i \frac{\pi}{\lambda z} ((m_2 \Delta x_2)^2+(n_2 \Delta y_2)^2)) \\ 
& ~~~~~~~~ \fft{u'(m_1, n_1)}
\end{aligned}
\label{eqn:eqn:fre_fourier3}
\end{equation}

Note that the sampling spacing on the destination plane is scaled to $\Delta x_2={\Delta x_1}/(\lambda z)$ and $\Delta y_2 = {\Delta y_1}/(\lambda z)$.

\subsubsection{Shifted Fresnel diffraction}

Shifted-Fresnel diffraction \cite{shift_fre1} enables arbitrary sampling pitches to be set on the source and destination planes as well as a shift away from the propagation axis, which is referred as to off-axis propagation.
The equation is derived from a combination of Fourier form Fresnel diffraction and scaled Fourier transform \cite{scaled_fourier}.
The same methods were proposed in somewhat different areas \cite{shift_fre2, shift_fre3}.
The discretized shifted Fresnel diffraction is expressed by the following equations:
\begin{equation}
\begin{aligned}
u_2[m_2,n_2] = C_S \ifft{ \fft{u_1'[m_1,n_1]} \fft{h_S[m_1, n_1]} } 
\end{aligned}
\label{eqn:eqn:shift_fre1}
\end{equation}
\begin{equation}
\begin{aligned}
C_S & = \frac{\exp(i k z)}{i \lambda z} \exp(i \frac{\pi}{\lambda z} (x_1^2+y_1^2)) \\
& \exp(-i \frac{2 \pi}{\lambda z}(O_{x_0} p_{x_0} x_1 +  O_{y_0} p_{y_0} y_1  ))  \exp(- i \pi(S_x m_1^2 + S_y n_1^2))
\end{aligned}
\label{eqn:eqn:shift_fre2}
\end{equation}
\begin{equation}
\begin{aligned}
u_1'[m_1,n_1] & = u_1[m_1,n_1] \exp(i \frac{\pi}{\lambda z}(x_0^2+y_0^2)) \\
& \exp(-i 2 \pi (m_0 S_x O_{x_1}  + n_0 S_y O_{y_1} )) \\
& \exp(- i \pi (S_x m_0^2 + S_y n_0^2 ))
\end{aligned}
\label{eqn:eqn:shift_fre3}
\end{equation}

\begin{equation}
h[m,n] = \exp(i \pi( S_x m^2 + S_y n^2 )) 
\end{equation}
where, $(O_{x_0}$, $O_{y_0})$ and $(O_{x_1}$, $O_{y_1})$ are the shift distances away from the propagation axis and $S_x, S_y, x_0, y_0, x_1, y_1$ are defined as follows:
\begin{equation}
\begin{aligned}
& S_x  =\frac{p_{x_0} p_{x_1}}{\lambda z}, ~~~
  S_y  =\frac{p_{y_0} p_{y_1}}{\lambda z}, \\
& x_0  =m_0 p_{x_0} + O_{x_0}, ~~~
y_0  =n_0 p_{y_0} + O_{y_0}, \\
& x_1  =m_1 p_{x_1} + O_{x_1},   ~~~
y_1  =n_1 p_{y_1} + O_{y_1}. 
\end{aligned}
\end{equation}
For more details, see Ref. \cite{shift_fre1}.

\subsubsection{Angular spectrum method}

The angular spectrum method can be devised from the Helm-holtz equation and is 
suitable for computing diffraction at short distance, which is impossible for Fresnel and Shifted Fresnel diffractions.
The angular spectrum method is expressed by:

\begin{equation}
\begin{aligned}
u(x,y)  = \int \!\!\int_{-\infty}^{+\infty} & A(f_x, f_y, 0) H_{A}(f_x,f_y) \\
& ~~~~~ \exp(i \ 2 \pi (f_x x + f_y y)) d f_x d f_y,
\end{aligned}
\label{eqn:ang1}
\end{equation}
where, $f_x$ and $f_y$ are spatial frequencies, $A(f_x, f_y,0)$ is defined by $A(f_x, f_y,0)=\mathcal{F}[u(x_1, y_1)]$, and $H_{A}(f_x, f_y)$ is the transfer function,
\begin{equation}
H_{A}(f_x,f_y)=\exp(i z \sqrt{k^2 - 4 \pi^2 (f_x^2 + f_y^2)})  
\label{eqn:ang2}
\end{equation}
where, $k=2 \pi / \lambda$ is the wave-number.

The discretizing frequencies are $(f_x, f_y)=(m_1 \Delta f_x, n_1 \Delta f_y)$, where $\Delta f_x$ and $\Delta f_y$ are the sampling pitches on the frequency domain.
Therefore, the discretized angular spectrum method is expressed by:
\begin{equation}
u(m,n) = \ifft{
\fft{u(m,n)} H_A(m_1,n_1)}.
\label{eqn:ang3}
\end{equation}

\subsubsection{Shifted angular spectrum method}

The angular spectrum method calculates the exact solution in scalar light propagation because it is derived from the Helmholtz equation without using any approximation, but only calculates light propagation when close to a source plane due to the aliasing problem.
In addition, the angular spectrum method does not calculate off-axis propagation.

The shifted angular spectrum method \cite{shift_angular} method enables off-axis propagation by applying a band-limited function to Eq. (\ref{eqn:ang2}).
In addition, although the angular spectrum method, as mentioned, triggers an aliasing error when calculating a long propagation distance \cite{band_angular}, the shifted angular spectrum method overcomes the aliasing problem, making it an excellent means of performing diffraction calculations.
The discretized shifted angular spectrum method is expressed by:
\begin{equation}
\begin{aligned}
u(m_2,n_2) & =  \ifft{ \fft{u(m_1,n_1)} H_{SA}(m_1,n_1) \\ 
& Rect(\frac{m_1 - c_x}{w_x}) Rect(\frac{n_1 - c_y}{w_y})
}
\end{aligned}
\label{eqn:shift_ang1}
\end{equation}
where, $H_{SA}$ is the shifted transfer function of the same,
\begin{equation}
\begin{aligned}
H_{SA}(m_1,n_1) = H_{A}(m_1,n_1) \exp(2 \pi (O_x m_1 \Delta f_x + O_y n_1 \Delta f_y) )
\end{aligned}
\label{eqn:shift_ang2}
\end{equation}

Two rectangle functions are band-limit functions with horizontal and vertical band-widths of $w_x, w_y$ and shift amounts of $c_x, c_y$.
See Ref. \cite{shift_angular} for the determination of these parameters.

\subsubsection{Summary of diffraction calculation}

As mentioned, many numerical diffraction calculations have been proposed to date, a classification of which is shown in Table \ref{tbl:diffraction}.
Each diffraction must be used in terms of the number of FFTs (namely, which is proportional to the calculation time), required memory, propagation distance, sampling pitches and on- or off-axis propagation.
For example, if we calculate the diffraction calculation with different sampling pitches on source and destination planes, we must use Shifted Fresnel diffraction.

Note that, in the diffraction calculations of the convolution form, the area of the source and destination planes must be expanded from $N_x \times N_y$ to $2 N_x \times 2 N_y$ during the calculation, because we avoid aliasing by circular convolution.
After the calculation, we extract the exact diffracted area on the destination plane with the size of $N_x \times N_y$.
Therefore, we need FFTs and memories of size $2 N_x \times 2 N_y$ during the calculation.

\begin{table*}
\caption{Classification of diffraction calculation. In the diffraction calculations of the convolution form, the area of the source and destination planes must be expanded from $N_x \times N_y$ to $2 N_x \times 2 N_y$ during the calculation, because we avoid aliasing by circular convolution.
After the calculation, we extract the exact diffracted area on the destination plane with the size of $N_x \times N_y$.
Therefore, we need FFTs and memories of size $2 N_x \times 2 N_y$ during the calculation.}
\begin{center}
\begin{tabular}{|c|c|c|c|c|c|c|}
\hline
Diffraction & \shortstack{Propagation \\ distance} & \shortstack{Sampling \\ pitch on \\ source \\ plane} & \shortstack{Sampling \\ pitch on \\ destination \\ plane} & \shortstack{Off-axis \\ calculation} & \shortstack{Number \\ of \\  FFTs} & \shortstack{Required \\ memory} \\ \hline
\shortstack{Fresnel \\ diffraction \\ (Convolution form)} & Fresnel & $\Delta x \times \Delta y$ & $\Delta x \times \Delta y$ & N.A. & 3 & $2N \times 2N$ \\ \hline
\shortstack{Fresnel \\ diffraction \\ (Fourier form)} & Fresnel & $\Delta x \times \Delta y$ & $\frac{\Delta x}{\lambda z} \times \frac{\Delta y}{\lambda z}$ & N.A. & 1 & $N \times N$ \\ \hline
\shortstack{Shifted \\ Fresnel \\ diffraction} & Fresnel & Arbitrary & Arbitrary & Available & 3 & $2N \times 2N$ \\ \hline
\shortstack{Angular \\ spectrum \\ method} & Short distance & $\Delta x \times \Delta y$ & $\Delta x \times \Delta y$ & N.A. & 2 & $2N \times 2N$ \\ \hline
\shortstack{Shifted \\ angular \\ spectrum method} & All & $\Delta x \times \Delta y$ & $\Delta x \times \Delta y$ & Available & 2 & $2N \times 2N$ \\ \hline
\end{tabular}
\end{center}
\label{tbl:diffraction}
\end{table*}

\subsection{CWO class: Simple example using the CWO++ library of the Fresnel diffraction calculation on the CPU}
\label{CWO}

The CWO class provides diffraction calculations and auxiliary functions.
We used the FFTW library \cite{fftw} in the diffraction calculations in the CWO class.

We show the following source-code of the Fresnel diffraction calculation on a CPU using the CWO++ library:

\begin{lstlisting}[caption={Fresnel diffraction calculation on a CPU using the CWO class.}, label=ex1, language=C, numbers=left, numberstyle=\tiny, stepnumber=1, frame=single]
CWO c; 
c.Load("lena512x512.bmp");
c.Diffract(0.2, CWO_FRESNEL_CONV);
c.Intensity();
c.Scale(255);
c.Save("lena512x512_diffract.bmp");
\end{lstlisting}

\begin{figure}[htb]
\centerline{
\includegraphics[width=9cm]{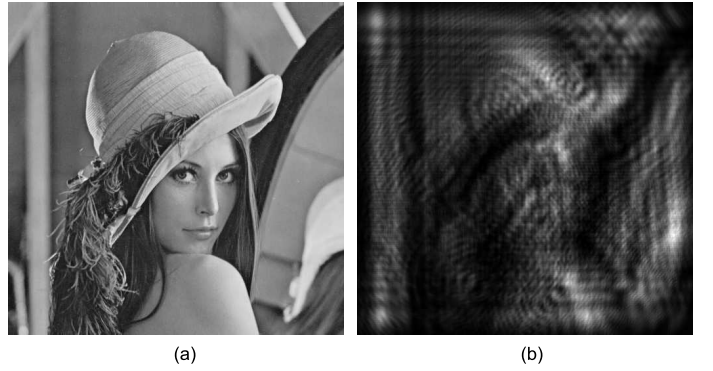}}
\caption{(a) Original image with $512 \times 512$ pixels. (b) Diffracted image with the wavelength of 633 nm and the sampling pitches of $10 \mu m \times 10 \mu m$ on the source and destination planes.}
\label{fig:lena-diff}
\end{figure}

In line 1, we define the instance ``c" of the CWO class.
In the next line, using the member function of ``Load" in the CWO class, we store an original image (Fig.\ref{fig:lena-diff}(a)) of a bitmap file ``lena512
x512.bmp" with $512 \times 512$ pixels into the instance ``c". 
``Load" can also read jpeg tiff formats, and so forth. 
See more details in  \ref{formats}.
In line 3, the CWO member function ``Diffract" calculates the diffraction calculation according to the first and second arguments of the function, which are the propagation distance and type of diffraction calculation, respectively. 
Here, we select the propagation distance of 0.2 m and Fresnel diffraction of the convolution form (CWO\_FRESNEL\_CONV).
The calculation results in a complex amplitude field.

To observe the light intensity field of the complex amplitude field as image data, the initial step involves the CWO member function ``Intensity" calculating the absolute square of the complex amplitude field in line 4.
Namely, the calculation is expressed by:
\begin{equation}
I(m_2,n_2)=|u_2(m_2,n_2)|^2 .
\end{equation}

In the next, the CWO member function ``Scale" converts the intensity field with a large dynamic range into that with 256 steps.
\begin{equation}
I_{256}(m_2,n_2)=\frac{I(m_2,n_2) - min\{I(m_2,n_2)\}}{max\{I(m_2,n_2)\}-min\{I(m_2,n_2)\}} \times 255,
\label{eqn:norm}
\end{equation}
where, the operators $max\{\cdot\}$ and $min\{\cdot\}$ take the maximum and minimum values in the argument.
Finally, we save the intensity field with 256 steps as a bitmap file ``lena512x512\_diffract.bmp".
Of course, we can also save the intensity field in other image formats.
Figure \ref{fig:lena-diff} (b) shows the diffracted image.

Note that the sample code does not explicitly indicate the wavelength and sampling pitches on the source and destination planes.
The default wavelength and sampling rates are 633 nm and $10 \mu m \times 10 \mu m$.
If we change the wavelength and sampling pitch, we can use the CWO member functions ``SetWaveLength" and ``SetPitch".

\subsection{Three-dimensional diffraction calculations}

\begin{figure}[htb]
\centerline{
\includegraphics[width=8cm]{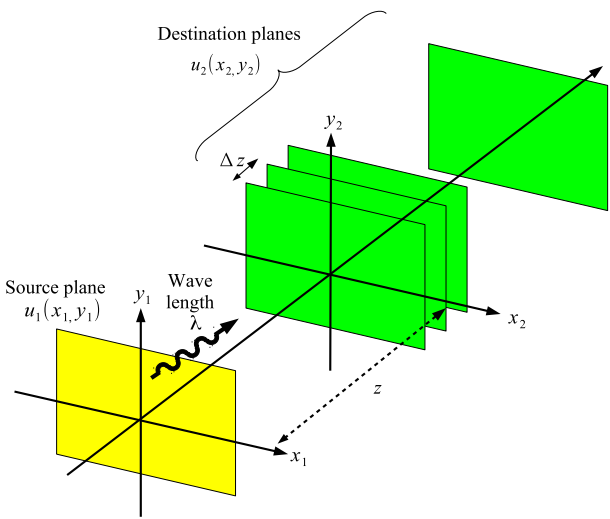}}
\caption{Scalar 3D diffraction field by stacking 2D diffraction calculations.}
\label{fig:3d_diff}
\end{figure}

Certain methods for calculating a scalar 3D field were proposed \cite{3d,3d_direct}.
In Ref. \cite{3d_direct}, the method can calculate a 3D diffraction calculation in a low numerical aperture (NA) system using once 3D-FFT, so that the method is effective in terms of computational cost and memory amount.
In the CWO++ library, we provide a calculation of a scalar 3D diffraction field by stacking 2D diffraction calculations as mentioned in a depth direction \cite{3d}.
Figure \ref{fig:3d_diff} shows a scalar 3D diffraction field by stacking 2D diffraction calculations.
Unlike 2D diffraction calculations, the 3D diffraction calculation requires a sampling pitch $\Delta z$ in a depth direction.

List \ref{stack_3d} shows the 3D diffraction calculation from a hologram, on which two small circles are recorded by the angular spectrum method.
In the calculation, the conditions are as follows: the number of pixels in 3D diffracted field is $512 \times 512 \times 512$, the horizontal and vertical sampling pitches are $10 \mu m \times 10 \mu m$, the sampling pitch in a depth direction $\Delta z$ is $1 cm$, the distance between the two small circles and the hologram is $10 cm$ and the radii of the two small circles are $270 \mu m$.
We need to set the large sampling pitch in a depth direction compared with the horizontal and vertical sampling pitches because the optical system has a low NA.

\begin{figure}[htb]
\centerline{
\includegraphics[width=8cm]{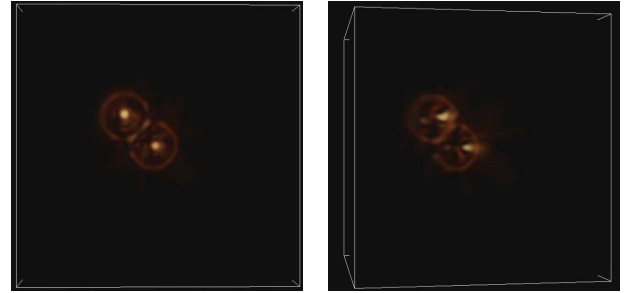}}
\caption{Volume rendering of the 3D diffraction calculation by List \ref{stack_3d}.}
\label{fig:volume}
\end{figure}

In line 1, we prepare two instances ``c1" and ``c2" because we need to maintain a hologram and a 3D diffracted field, individually.
In line 2, we load the hologram to the instance c1.
In line 3, we allocate the memory of $512 \times 512 \times 512$ pixels for the 3D diffraction field.
In line 4, we set horizontal, vertical and depth-direction sampling pitches, respectively.
In line 5, the 3D diffraction field is calculated by the member function ``Diffract3D".
The first argument is set to the instance of the CWO class maintaining the hologram.
The second argument is the distance $z$ as shown in Fig. \ref{fig:3d_diff} between the hologram and 3D diffracted field.
The third argument concerns the type of diffraction calculation.
In line 6, we save the calculated 3D diffraction field as a cwo file, which includes the 3D diffraction field in complex amplitude in binary form.
Figure \ref{fig:volume} shows the volume rendering from the cwo file.
In the left figure, we can see the two circles reconstructed head-on in the volume.
In the right figure, we can see a different view of the reconstructed two circles.

\begin{lstlisting}[caption={3D diffraction calculation based on the stack of 2D diffracted planes.}, label=stack_3d, language=C, numbers=left, numberstyle=\tiny, stepnumber=1, frame=single]
CWO c1,c2;
c1.Load("hologram.bmp");
c2.Create(512, 512, 512);
c2.SetPitch(10e-6, 10e-6, 0.01);
c2.Diffract3D(a1, -0.1, CWO_ANGULAR);
c2.Save("3d.cwo");
\end{lstlisting}

\subsection{Calculations of a complex amplitude field and a computer generated hologram from point light sources}
\label{cwoPLS}

Current CGH calculations are mainly categorized into two approaches: polygon-based \cite{ExtreamCGH, Ahrenberg} and Point Light Sources (PLSs) approaches \cite{nlut, image_hol, lut}. 
Polygon-based approaches are based on diffraction calculations using FFTs.
In the subsection, we focus on the PLS approach, which treats a 3D object as a composite of multiple PLSs, and generates a CGH from a 3D object more flexibly than polygon-based approaches.

\begin{figure}[htb]
\centerline{
\includegraphics[width=8cm]{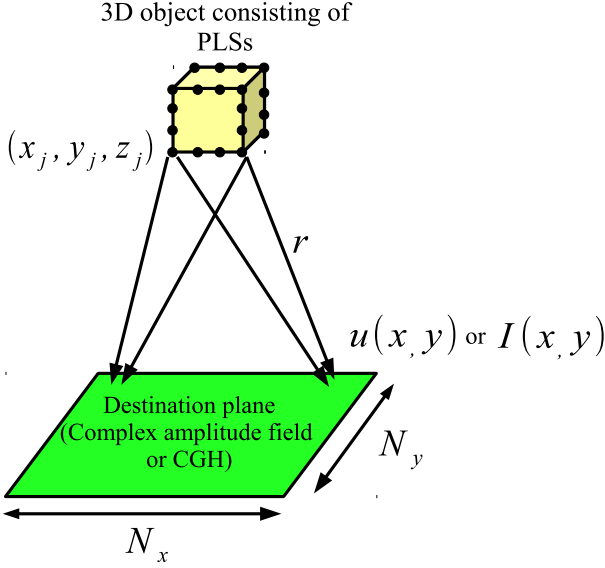}}
\caption{PLS-based diffraction and CGH calculations.}
\label{fig:pls}
\end{figure}

The class ``cwoPLS" calculates a complex amplitude field and CGH from PLSs.
In Fig. \ref{fig:pls}, the calculations assume that a 3D object is composed of PLSs with a number of $N$.

We can calculate the complex amplitude field or CGH to superimpose PLS distributions from one PLS to the destination plane. 
The complex amplitude field $u(x,y)$ from PLSs by Fresnel diffraction is expressed as,

\begin{equation}
\begin{aligned}
u(x,y)=\frac{\exp(i \frac{2 \pi}{\lambda} z)}{i \lambda z} \sum_j^N A_j {\exp}(\frac{2 \pi}{\lambda} (\frac{(x-x_j)^2+(y-y_j)^2}{2 z_j}))
\end{aligned}
\label{eqn:pls_fre}
\end{equation}
where, $(x,y)$ and $(x_j, y_j, z_j)$ are coordinates on the destination plane and a 3D object, $A_j$ is the light intensity of the 3D object.
A CGH calculation based on Fresnel diffraction is expressed as \cite{lut},
\begin{equation}
\begin{aligned}
I(x,y) & =\sum_j^N {A_j}{\rm cos}(\frac{2 \pi}{\lambda} (\frac{(x-x_j)^2+(y-y_j)^2}{2 z_j}))
\end{aligned}
\label{eqn:pls_fre_cgh}
\end{equation}
where, $I(x,y)$ is the CGH pattern.
The computational complexity of the above formulas are O($N N_x N_y$), where $N_x$ and $N_y$ are the horizontal and vertical pixel numbers of the CGH.

List \ref{lst:plscgh} shows CGH generation using the class ``cwoPLS".
We generate a $2,048 \times 2,048$ pixels CGH from the 3D object of a dinosaur with 11,646 points.
The sampling pitch on the CGH, wavelength (default value) and distance between the CGH and the 3D object is $4 \mu m \times 4 \mu m$, $633 nm$ and 0.2 m, respectively.
The class treats the 3D object as the source and the CGH as the destination plane.

In line 4 of List \ref{lst:plscgh}, the member function ``Create" allocates the required memory of the CGH.
In line 5, the member function ``SetDstPitch" is set to the sampling pitches on the destination plane, namely the CGH.
In line 6, the first and second arguments of the member function ``SetOffset" are set to the offsets of the 3D object, namely horizontal and vertical offsets of 1 cm (2500 pixels) and -2 mm (-512 pixels) from the origin, respectively.
In addition, the third argument sets the distance between CGH and a 3D object of 0.2 m.
In line 7, the 3D object data of the dinosaur is read, while in the next line, the coordinates of the 3D object are normalized from -4 mm (-1000 pixels) to 4 mm (1000 pixels).
In line 9, the CGH is calculated by Eq. (\ref{eqn:pls_fre_cgh}), and in the next line, the calculated CGH pattern is normalized to 256 steps by Eq. (\ref{eqn:norm}).

Figure \ref{fig:pls}(a) and (b) show the CGH generated by List.\ref{lst:plscgh} and the reconstructed image from (a) using the CWO class.

\begin{lstlisting}[caption={CGH generation using the class ``cwoPLS".}, label=lst:plscgh, language=C, numbers=left, numberstyle=\tiny, stepnumber=1, frame=single]
cwoPLS c;
float p=4e-6;	
float z=0.2;
c.Create(2048,2048);
c.SetDstPitch(p,p);
c.SetOffset(2500*p, -512*p, z);
c.Load("tyranno_000.3df");
c.ScalePoint(p*1000);
c.PLS(CWO_PLS_FRESNEL_CGH);
c.Scale(255);
c.Save("cgh.bmp");
\end{lstlisting}

\begin{figure*}[htb]
\centerline{
\includegraphics[width=15cm]{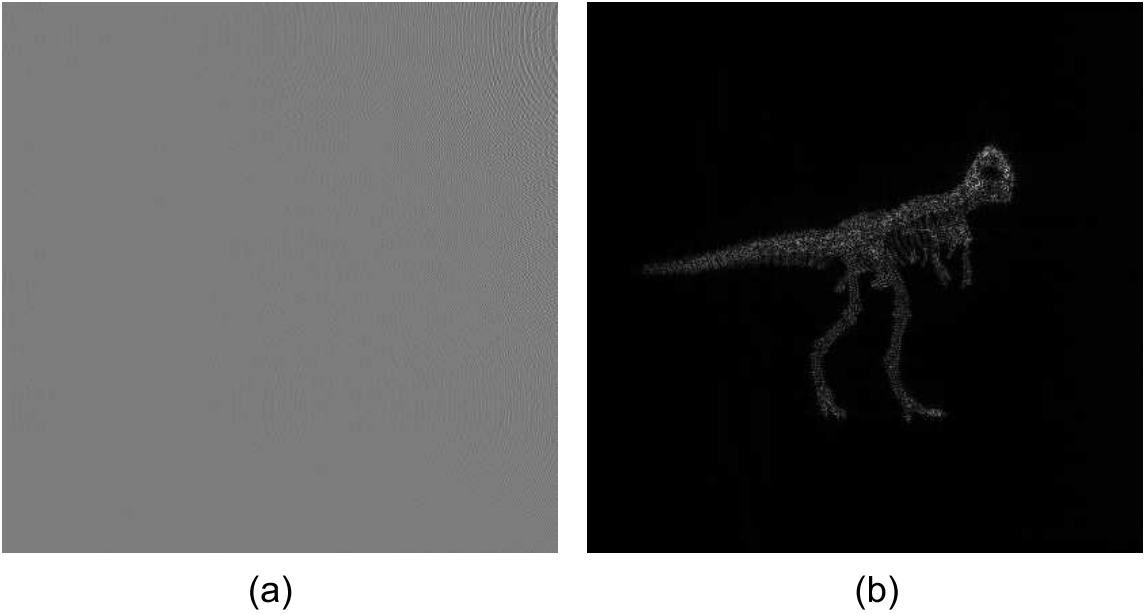}}
\caption{(a) CGH generated by the class cwoPLS. (b) Reconstructed image from (a) using the class CWO.}
\label{fig:pls_cgh}
\end{figure*}


\section{GWO and gwoPLS classes: Diffraction and CGH calculations on GPU}

The current CWO++ library provides diffraction and CGH calculations on NVIDIA GPU chips by the GWO and gwoPLS classes.
In this subsection, we briefly describe an NVIDIA GPU, and subsequently show the source code of Fresnel diffraction and CGH calculations on a GPU using the GWO and gwoP \\
LS classes.

Although initially GPU chips were mainly used for rendering 3D computer graphics, recent GPU chips have also been used for numerical computation.
In 2007, NVIDIA released a new GPU architecture and its software development environment, Compute Unified Device Architecture (CUDA).
CUDA can be used to facilitate the programming of numerical computations more than software previously developed, such as HLSL, Cg language and so forth.
Since its release, many papers using NVIDIA GPU and CUDA have been published in optics. 
In particular, calculations of CGH \cite{lucenteGPU, masuda3, Ahrenberg, Onural, lutgpu} and reconstruction in digital holography \cite{gwo, realtimeDHM, multiDHM, DHAhrenberg, DHCarl, DHNaughton, DHGarcia} have successfully accelerated these calculations.

\begin{figure}[htb]
\centerline{
\includegraphics[width=8cm]{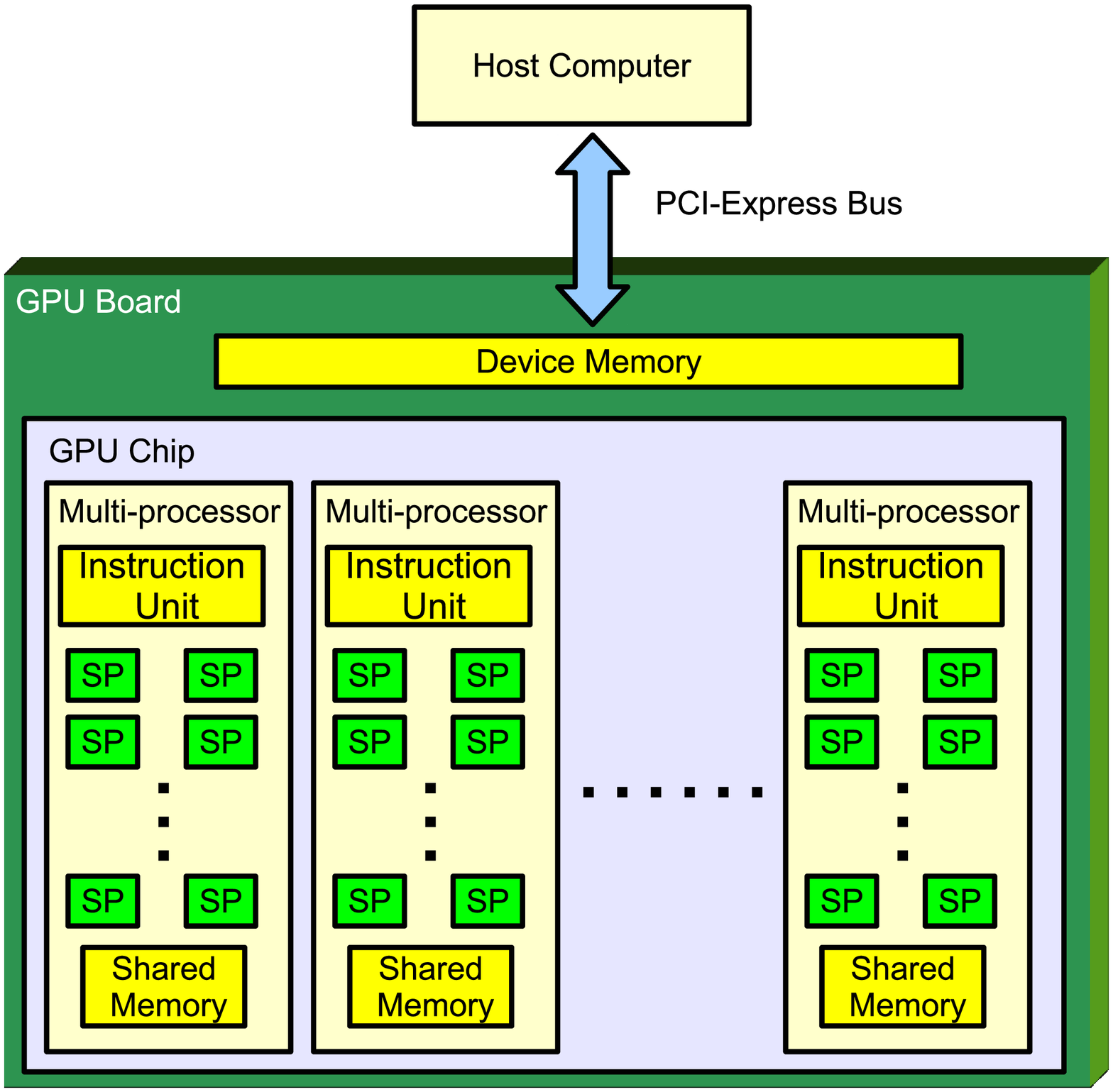}}
\caption{Structure of NVIDIA GPU chip.}
\label{fig:gpu}
\end{figure}

Figure \ref{fig:gpu} shows the structure of an NVIDIA GPU, featuring GPU chips with some Multi-Processors (MP).
Moreover, the MP includes Stream Processors (SP), where the number of SPs per MP differs depending on the version of the NVIDIA GPU chip.
One SP can operate 32- or 64-bit floating-point addition, multiplication and multiply-add instructions.
SPs in an MP operate the same instruction in parallel, and process different data.
Namely, an MP is a Single Instruction Multiple Data (SIMD) fashion processor.
In addition, each multiprocessor can operate the same or different processing, thus allowing the GPU chip to be used as a highly parallel processor. 
The specifications of the GPUs used in this paper are shown in Table \ref{tbl:gpu}.

\begin{table*}[htbp]
\caption{Specifications of the GPUs used in this paper.}
\begin{center}
\begin{tabular}{|l|r|r|r|}
\hline
 & \multicolumn{1}{l|}{NVIDIA Geforce GTX 460M} & \multicolumn{1}{l|}{GeForce GTX295 (1 chip)} & \multicolumn{1}{l|}{GeForce GTX 580} \\ \hline
CUDA processors & 192 & 240 & 512 \\ \hline
Memory amout (GBytes) & 1.5 & 0.896 & 1.53 \\ \hline
Core clock(GHz) & 1.35 & 1.24 & 1.54 \\ \hline
Memory clock (GHz) & 1.25 & 1 & 2 \\ \hline
Memory band width (GByte /sec) & 60 & 223.8 & 192.4 \\ \hline
\end{tabular}
\end{center}
\label{tbl:gpu}
\end{table*}

The host computer controls the GPU board and the communication between the host computer and the GPU board via the PCI-express bus.
The host computer can also directly access the device memory on the GPU board, which is used for storing input data, namely the location of the source plane or 3D object, and the destination plane as computed by the GPU.
The multiprocessor has a shared memory, which is limited, but low latency and faster than the device memory.
In the classes, these memories are used for fast calculation.

The CUDA compiler compiles a C-like language source code for the instruction set of the GPU chip, which is referred to as a ``Kernel".
We download this ``Kernel" to the GPU chip via the PCI-express bus.
The kernel codes in the CWO++ library are collected in the form of a library and DLL files on Windows OS, named ``gwo\_lib.lib" and ``gwo\_lib.dll".
When we use the GWO or gwoPLS classes, we need to link these files to our program.
See more details in \ref{appendix_install}.

The diffraction calculations, as mentioned, require some FFT operations.
The CUDA compiler allows us to accelerate the FFT algorithm on an NVIDIA GPU chip, which is named CUFFT.
It is similar to the FFTW library \cite{fftw}, and we use CUFFT for the GWO and gwoPLS classes.

When the GWO and gwoPLS classes are used, they implicitly execute the following steps, which are hidden to CWO++ library users: 

\begin{enumerate}
\item Allocating the required memories for calculation on the device memory.
\item Sending the input data (source plane or 3D object data) to the allocated device memory.
\item Executing kernel functions. The result (complex amplitude field or CGH on the destination plane) is stored in the allocated device memory.
\item Receiving the result from the allocated device memory.
\item Releasing the allocated device memories.
\end{enumerate}

\subsection{Fresnel diffraction and CGH calculations on a GPU using the GWO and gwoPLS classes}
We show the source-code of a Fresnel diffraction calculation on a GPU using the GWO class. 
This example is the same as Section \ref{CWO} except using GPU, but somewhat different in comparison to List \ref{ex1}.

\begin{lstlisting}[caption={Fresnel diffraction calculation on a GPU using class GWO.}, label=gpu_fre, language=C, numbers=left, numberstyle=\tiny, stepnumber=1, frame=single]
CWO c;
GWO g; 
c.Load("lena512x512.bmp");
g.Send(c);
g.Diffract(0.2, CWO_FRESNEL_CONV);
g.Intensity();
g.Scale(255);
g.Recv(c);
c.Save("lena512x512_diffract.bmp");
\end{lstlisting}

As the initial change, we define the instance ``g" of the GWO class, in line 2.
As the second change, the source plane, namely Fig.\ref{fig:lena-diff}(a), is sent to a GPU by using the GWO class member function ``Send".
When calling the member function ``Send", the first step involves the automatic allocation of the required device memory on the GPU to the instance ``g", via the CUDA API function ``cudaMalloc".
Subsequently, the source plane on the host memory is sent to the device memory on the GPU using the CUDA API function ``cudaMemcpy".

In lines 5 to 7 of List \ref{gpu_fre}, we change the instance names in lines 3 to 5 of List \ref{ex1} from the instance ``c" of the CWO class to the instance ``g" of the GWO class.
Therefore, these functions calculate the diffraction, intensity and normalization on the GPU. Finally, we receive the calculation result from the GPU to the host in line 8.

Next, we show the source-code of CGH calculation on a GPU using the GWO class.
This example is the same as Section \ref{cwoPLS} except using GPU.
There are also some changes as compared with List \ref{lst:plscgh}.

\begin{lstlisting}[caption={CGH calculation on a GPU using class gwoPLS.}, label=gpu_cgh, language=C, numbers=left, numberstyle=\tiny, stepnumber=1, frame=single]
cwoPLS c;
gwoPLS g;
float p=4e-6;	
float z=0.2;
c.Create(2048,2048);
c.SetDstPitch(p,p);
c.SetOffset(2500*p, -512*p, z);
c.Load("tyranno_000.3df");
g.ScalePoint(p*1000);
g.Send(c);
g.PLS(CWO_PLS_FRESNEL_CGH);
g.Scale(255);
g.Recv(c);
c.Save("cgh.bmp");
\end{lstlisting}

\section{Implementations of the CWO and GWO classes}

Now, we describe the implementations of the CWO and GWO classes.
For simplicity, we select the implementations of the Fresnel diffraction (Eq. (\ref{eqn:fre_conv_fft})) on the CWO and GWO classes as an example.

\begin{figure*}[htb]
\centerline{
\includegraphics[width=15cm]{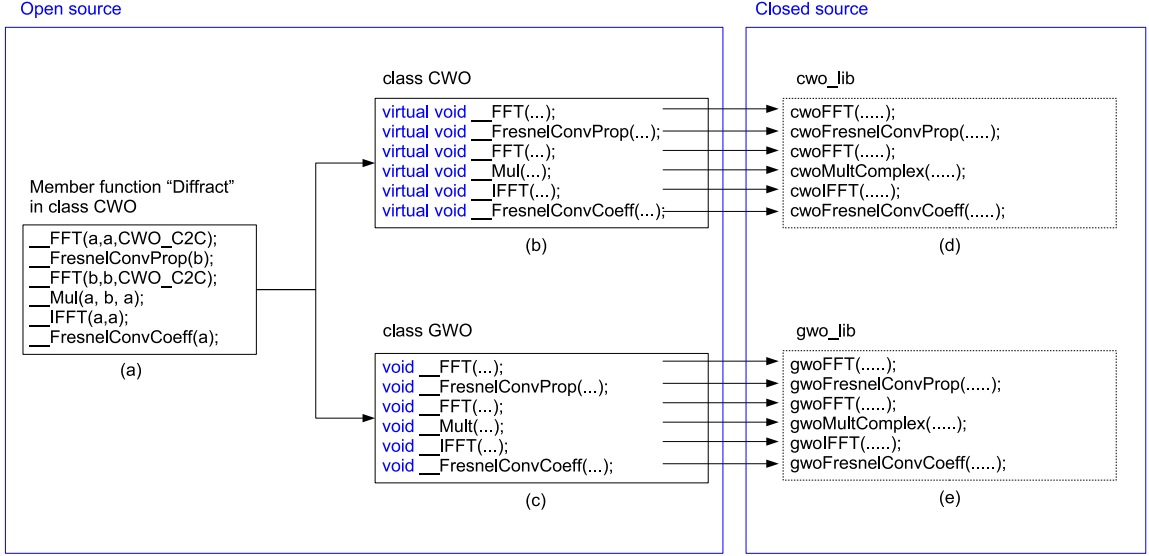}}
\caption{Implementations of Fesnel diffraction in CWO and GWO. }
\label{fig:implement}
\end{figure*}

Figure \ref{fig:implement} shows the implementations.
Figure \ref{fig:implement}(a) shows a portion of the CWO member function ``Diffract", the functions in which are defined as virtual functions of C++.
The functions ``FFT", ``FresnelConvProp", ``Mult", ``IFFT" and ``FresnelConvCoeff" calculate FFT, the impulse response of Eq. (\ref{eqn:fre_conv_prop}), complex multiplication, inverse FFT and the coefficient of Eq. (\ref{eqn:fre_conv_fft}).

When the function ``Diffract" in the CWO class is called, the virtual functions defined in the CWO class are called (Fig. \ref{fig:implement}(b)).
Moreover, these functions also call the functions defined in the library and DLL files of ``cwo\_lib" (Fig. \ref{fig:implement}(d)), which are activated on a CPU.

Conversely, when the function ``Diffract" in the GWO class is called, the virtual functions defined in the GWO class are called (Fig.\ref{fig:implement}(c)).
Moreover, these functions call those defined in the library and DLL files of ``gwo\_lib" (Fig.\ref{fig:implement}(e)).
These functions are activated on a GPU.
Therefore, the function ``Diffract" in the GWO class calculates the Fresnel diffraction on a GPU.

The classes shown in the Table \ref{tbl:cwo} are distributed as open-source code, while the functions in the library and DLL files of ``cwo\_lib" and ``gwo\_lib", which are closed source code, are distributed as binary files.

The current CWO++ library is compatible with Intel and AMD x86 CPUs and NVIDIA GPUs. 
If we port the CWO++ library to other hardware, we need to inherit the CWO class to a new class indicating the new hardware, then re-define virtual functions and write functions corresponding to ``gwo\_lib".

For instance, new GPUs of the HD5000 and HD6000 series made by AMD were released. These GPUs have new architecture and software environment, Open Computing Language (OpenCL). The architectures also differ from NVIDIA GPUs. 
We have already reported the CGH calculation of Eq. (\ref{eqn:pls_fre_cgh}) \cite{amd_cgh} and Fresnel diffraction calculation \cite{amd_diff}.
Although we do not implement the CWO++ library on the AMD GPUs, we will make a new class for AMD GPUs using the above mentioned method.

\section{Field type}

The classes in the CWO++ library, namely the CWO, GWO, cwoPLS and gwoPLS classes, have field types indicating current fields.
There are three field types: complex amplitude field (the predefined macro: CWO\_FLD\_COMPLEX ), intensity field (the predefined macro: CWO\_FLD\_INTENSITY ) and phase field (the predefined macro: CWO\_FLD\_PHASE).

The above classes hold the field $u(x,y)$ for the source or destination plane.
If the field type is CWO\_FLD\_COMPLEX, the field $u(x,y)$ maintains a complex amplitude field as a complex number array,
\begin{equation}
\begin{aligned}
u(x,y)={\rm re}(x,y) + i ~ {\rm im}(x,y)
\end{aligned}
\label{eqn:field_complex}
\end{equation}
where ${\rm re}(x,y)$ and ${\rm im}(x,y)$ indicate real and imaginary components of the complex value on $(x,y)$.
Because the current CWO++ library has single floating point precisions for the real and imaginary components, the memory amount for the field $u(x,y)$ is $8 N_x N_y$ bytes where $N_x $ and $N_y$ are the number of pixels in the field.

If the field type is CWO\_FLD\_PHASE, the class maintains the phase field $\theta(x,y)$ as a real number array,
\begin{equation}
\begin{aligned}
\theta(x,y) = \tan^{-1} \frac{{\rm im}(x,y)}{{\rm re}(x,y)}
\end{aligned}
\label{eqn:field_phase}
\end{equation}

If the field type is CWO\_FLD\_INTENSITY, the class maintains a real number array $a(x,y)$ except the phase field.
The real number arrays include, for example, image data, amplitude, real or imaginary, only part of the complex amplitude field and so on.
The memory required for the intensity and phase fields is $4 N_x N_y$ bytes.

\begin{figure}[htb]
\centerline{
\includegraphics[width=9cm]{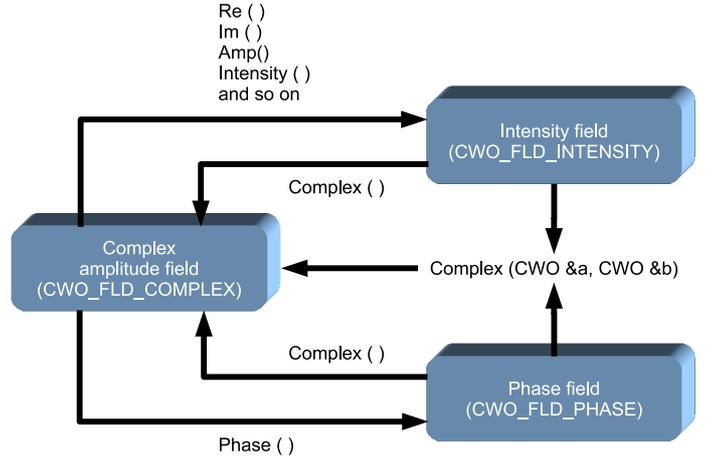}}
\caption{Field types and mutual conversions between each field.}
\label{fig:cwo_fld}
\end{figure}

Figure \ref{fig:cwo_fld} shows each field type and mutual conversions between each field.
We briefly describe the mutual conversions as follows:
\begin{enumerate}
\item If we use the member functions ``Re()" when the current field type is a complex amplitude field, the field is converted from this to a real part only and the field type is set to CWO\_FLD\_INTENSITY.

\item If we use member functions ``Phase()" when the current field type is a complex amplitude field, the field is converted from this to an argument of the same and the field type is set to CWO\_FLD\_PHASE.

\item If we use member functions ``Complex()" when the current field type is an intensity field, the field is converted from this to a complex amplitude field according to $u(x,y)\\
={\rm re}(x,y) + i ~ {\rm im}(x,y)$, where ${\rm re}(x,y)=a(x,y)$ is the intensity field and ${\rm im}(x,y)$ is zero and the field type is set to CWO\_FLD\_COMPLEX.

\item If we use member functions ``Complex()" when the current field type is a phase field, the field is converted from this to a complex amplitude field according to $u(x,y)={\rm re}(x,y) + i ~ {\rm im}(x,y)$, where ${\rm re}(x,y)=\cos(\theta(x,y))$ and ${\rm im}(x,y)\\
=\sin(\theta(x,y))$ and the field type is set to CWO\_FLD\_COM\\
PLEX.

\item If we use member functions ``Complex(CWO \&a, CWO \&b)" when the current field types of classes ``a" and ``b" are the intensity and phase fields respectively, the fields are converted to a complex amplitude field according to $u(x,y)=a(x,y) {\rm cos}(\theta(x,y)) + i ~ a(x,y) {\rm sin}(\theta(x,y))$, where ``a" and ``b" hold $a(x,y)$ and $\theta(x,y)$, respectively and the field type is set to CWO\_FLD\_COMPLEX.

\item If we use member functions ``Complex(CWO \&a, CWO \&b)" when the current field types of classes ``a" and ``b" are both intensity fields, the fields are converted to a complex amplitude field according to $u(x,y)={\rm re}(x,y) \\
+ i ~ {\rm im}(x,y)$, where ${\rm re}(x,y)$ and ${\rm im}(x,y)$ are the fields of ``a" and ``b", respectively and the field type is set to CWO\_FLD\\
\_COMPLEX.
\end{enumerate}

\begin{figure*}[htb]
\centerline{
\includegraphics[width=12cm]{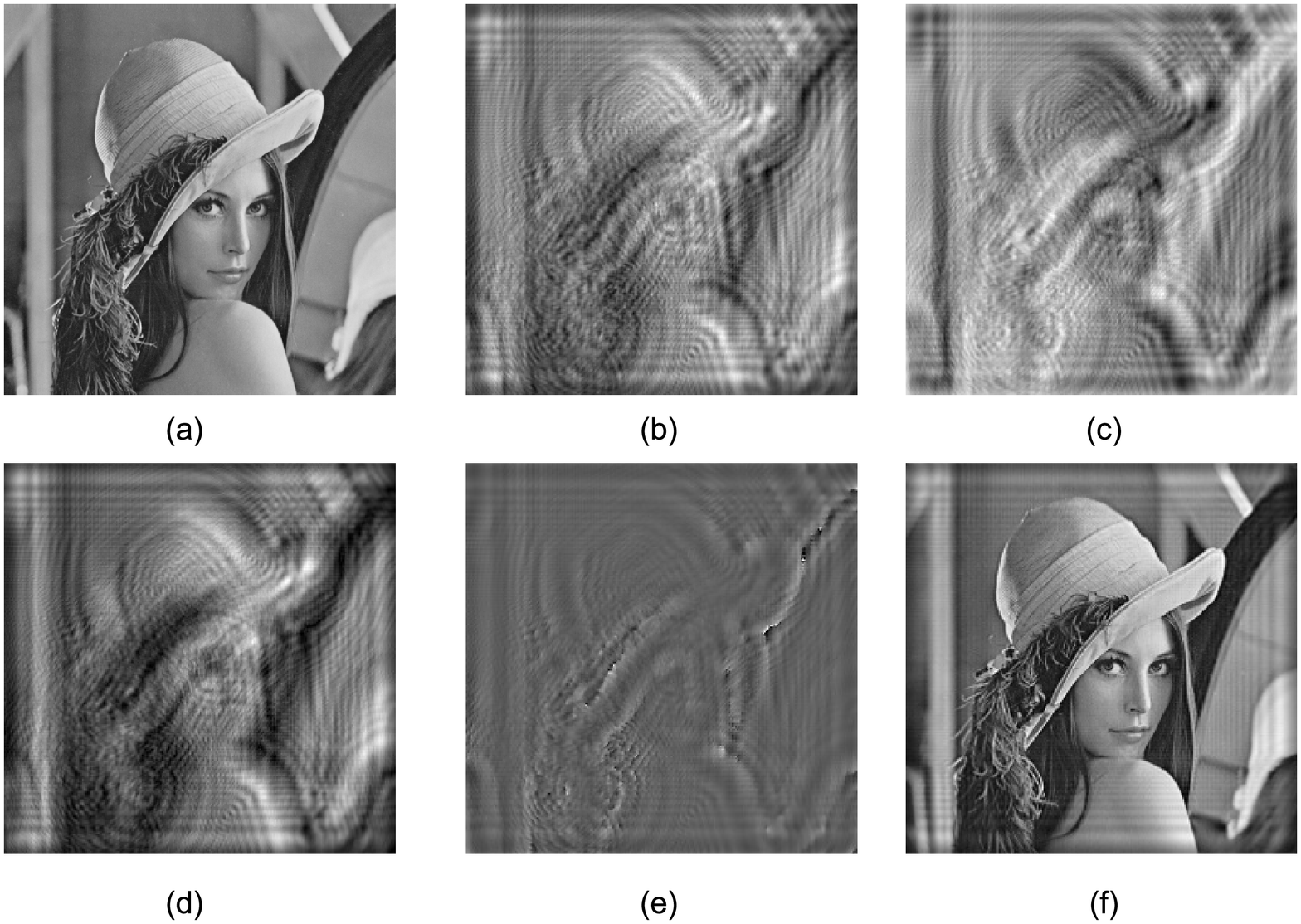}}
\caption{Resluts of mutual conversions between each field. (a) original image (b) real part of the diffracted light (c) imaginary part of the diffracted light (d) amplitude of the diffracted light (e) phase of the diffracted light and (f) back propagation from the complex amplitude of (d) and (e).}
\label{fig:conversion}
\end{figure*}

List \ref{lst:conversion} and Fig.\ref{fig:conversion} show examples of mutual conversions between each field type and their results, respectively. Figure \ref{fig:conversion} (a) is an original image. We calculate the diffracted light of the figure in lines 2 and 3 of List \ref{lst:conversion} because we observe the real part, imaginary part, amplitude and phase of the diffracted field respectively.

The real part of the diffracted light (Fig.\ref{fig:conversion}(b)) is obtained in lines 8 to 10, and, the imaginary part of the diffracted light (Fig. \ref{fig:conversion}(c)) is obtained in lines 11 to 13.
The amplitude
\begin{equation*}
\sqrt{{\rm re}(x,y)^2 + {\rm im}(x,y)^2} 
\end{equation*}
 of the diffracted light (Fig.\ref{fig:conversion}(d)) is obtained in lines 17 to 19, and, the phase of the diffracted light (Fig. \ref{fig:conversion}(e)) is obtained in lines 20 to 22.
In lines 26 to 28, we show the generation of the complex amplitude field from the instances ``a" and ``b", which hold amplitude (CWO\_FLD\_INTENSITY) and phase (CWO\_FL\\
D\_PHASE) respectively.
In lines 29 to 32, we show the result of the back propagation result (Fig.\ref{fig:conversion}(f)) from the position of the complex amplitude field to that of the original image.
In the result, although we observe a diffraction effect to some extent at the edges of the figure, the result is almost the same as the Fig.\ref{fig:conversion}(a).

\begin{lstlisting}[caption={Example of mutual conversions of field types.}, label=lst:conversion, language=C, numbers=left, numberstyle=\tiny, stepnumber=1, frame=single]
CWO a,b,c;
a.Load("lena512x512.bmp");
a.Diffract(0.1,CWO_ANGULAR);

b=a;
c=a;

b.Re();
b.Scale(255);
b.Save("re.bmp");
c.Im();
c.Scale(255);
c.Save("im.bmp");

b=a;
c=a;
b.Amp();
b.Scale(255);
b.Save("amp.bmp");
c.Phase();
c.Scale(255);
c.Save("phase.bmp");

b=a;
c=a;
b.Amp();
c.Phase();
a.Complex(b,c);
a.Diffract(-0.1,CWO_ANGULAR);
a.Re();
a.Scale(255);
a.Save("complex.bmp");
\end{lstlisting}

\section{Performance}

In this section, we show the calculation times of each diffraction calculation and CGH calculation on an Intel CPU and NVID\\
IA GPUs.
We used an Intel CPU, which was a Core i7 740 QM (with CPU clock frequency of 1.7GHz), and three GPUs, namely NVIDIA GeForce GTX 460M, GerForce GTX 295 and GeForce GTX580.
The GPU specifications are shown in Table \ref{tbl:gpu}.

In the diffraction calculations, the impulse response and transfer functions of each diffraction are only sufficient to calculate them once when the parameters, which are the propagation distance, the sampling pitches, wavelength, offsets and so on, are unchanged.
Of course, we need to re-calculate the impulse response and transfer functions when the parameters are changed.
Therefore, we evaluated the calculation times of each diffraction in both cases of re-calculation and once-calculation.

For the evaluation, we used Lists \ref{ex1} and \ref{gpu_fre} and changed the image size and diffraction type.
Table \ref{tbl:diff_time_recalc} shows the calculation times of diffraction calculations on the CPU and GPUs with recalculation of the impulse and transfer functions.
Table \ref{tbl:diff_time_once} shows the calculation times of diffraction calculations on the CPU and GPUs with the once-calculation of the impulse and transfer functions.
In the table, except for the diffraction type CWO\_FRESNEL\_FOURIER, we expand the area of the source and destination planes from $N_x \times N_y$ to $2 N_x \times 2 N_y$ during the calculation and avoid aliasing by circular convolution.
In the CPU calculations, we measured the time in line 3 of List \ref{ex1}.
In the GPU calculations, we measured the time in lines 4 and 5 of List \ref{gpu_fre}.

Table \ref{tbl:cgh_time} shows the calculation times of CGH calculations on the CPU and GPUs.
For the evaluation, we used Lists \ref{lst:plscgh} and \ref{gpu_cgh} and changed the number of PLSs.
In the CPU calculations, we measured the time in line 9 of List \ref{lst:plscgh}.
In the GPU calculations, we measured the time in lines 10 and 11 of List \ref{gpu_cgh}.

The calculation times of each diffraction calculation and CGH calculation on the GPUs were much faster than those of the CPU.

\begin{table*}[htbp]
\caption{Calculation times of each diffraction calculation on the CPU and GPUs with the re-calculation of the impulse and transfer functions.}
\begin{center}
\begin{tabular}{|c|c|c|c|c|}
\hline
\multicolumn{ 1}{|c|}{Resolution} & CPU (ms) & \multicolumn{ 3}{c|}{GPU (ms)} \\ \cline{ 2- 5}
\multicolumn{ 1}{|c|}{} & Intel Core i7 740QM & GeForce GTX 460M & GeForce GTX295 (1 chip) & GeForce GTX580 \\ \hline
\multicolumn{ 5}{|c|}{Fresnel diffraction convolution form (CWO\_FRESNEL\_CONV)} \\ \hline
$512 \times 512$ & 248 & 15 & 5 & 3 \\ \hline
$1024 \times 1024$ & $1.24 \times 10^3$ & 47 & 15 & 10 \\ \hline
$2048 \times 2048$ & $6.12 \times 10^3$ & 177 & 67 & 38 \\ \hline
\multicolumn{ 5}{|c|}{Fresnel diffraction Fourier form (CWO\_FRESNEL\_FRESNEL)} \\ \hline
$512 \times 512$ & 51.7 & 2.4 & 1 & 1 \\ \hline
$1024 \times 1024$ & 227 & 6 & 2 & 2 \\ \hline
$2048 \times 2048$ & 984 & 19 & 9 & 8 \\ \hline
\multicolumn{ 5}{|c|}{Shifted Fresnel diffraction (CWO\_SHIFTED\_FRESNEL)} \\ \hline
$512 \times 512$ & 477 & 15 & 5 & 3 \\ \hline
$1024 \times 1024$ & $2.07 \times 10^3$ & 48 & 16 & 10 \\ \hline
$2048 \times 2048$ & $9.48 \times 10^3$ & 186 & 71 & 40 \\ \hline
\multicolumn{ 5}{|c|}{Angular spectrum method (CWO\_ANGULAR)} \\ \hline
$512 \times 512$ & 260 & 12 & 3 & 2 \\ \hline
$1024 \times 1024$ & $1.17 \times 10^3$ & 36 & 11 & 8 \\ \hline
$2048 \times 2048$ & $5.56 \times 10^3$ & 135 & 47 & 29 \\ \hline
\multicolumn{ 5}{|c|}{Shifted angular spectrum method (CWO\_SHIFTED\_ANGULAR)} \\ \hline
$512 \times 512$ & 269 & 15 & 4 & 3 \\ \hline
$1024 \times 1024$ & $1.23 \times 10^3$ & 44 & 14 & 9 \\ \hline
$2048 \times 2048$ & $5.66 \times 10^3$ & 157 & 58 & 35 \\ \hline
\end{tabular}
\end{center}
\label{tbl:diff_time_recalc}
\end{table*}

\begin{table*}[htbp]
\caption{Calculation times of each diffraction calculation on the CPU and GPUs with the once-calculation of the impulse and transfer functions.}
\begin{center}
\begin{tabular}{|c|c|c|c|c|}
\hline
\multicolumn{ 1}{|c|}{Resolution} & CPU (ms) & \multicolumn{ 3}{c|}{GPU (ms)} \\ \cline{ 2- 5}
\multicolumn{ 1}{|c|}{} & Intel Core i7 740QM & GeForce GTX 460M & GeForce GTX295 (1 chip) & GeForce GTX580 \\ \hline
\multicolumn{ 5}{|c|}{Fresnel diffraction convolution form (CWO\_FRESNEL\_CONV)} \\ \hline
$512 \times 512$ & 117 & 10 & 3 & 2 \\ \hline
$1024 \times 1024$ & 620 & 29 & 11 & 6 \\ \hline
$2048 \times 2048$ & $3.30 \times 10^3$ & 104 & 48 & 26 \\ \hline
\multicolumn{ 5}{|c|}{Fresnel diffraction Fourier form (CWO\_FRESNEL\_FRESNEL)} \\ \hline
$512 \times 512$ & 52 & 2 & 1 & 1 \\ \hline
$1024 \times 1024$ & 229 & 5.4 & 2 & 2 \\ \hline
$2048 \times 2048$ & 993 & 19 & 9 & 8 \\ \hline
\multicolumn{ 5}{|c|}{Shifted Fresnel diffraction (CWO\_SHIFTED\_FRESNEL)} \\ \hline
$512 \times 512$ & 346 & 10 & 3 & 2 \\ \hline
$1024 \times 1024$ & $1.50\times 10^3$ & 30 & 11 & 7 \\ \hline
$2048 \times 2048$ & $6.91 \times 10^3$ & 110 & 51 & 28 \\ \hline
\multicolumn{ 5}{|c|}{Angular spectrum method (CWO\_ANGULAR)} \\ \hline
$512 \times 512$ & 121 & 9.5 & 3 & 2 \\ \hline
$1024 \times 1024$ & 624 & 26 & 10 & 6 \\ \hline
$2048 \times 2048$ & $3.30 \times 10^3$ & 97 & 44 & 24 \\ \hline
\multicolumn{ 5}{|c|}{Shifted angular spectrum method (CWO\_SHIFTED\_ANGULAR)} \\ \hline
$512 \times 512$ & 128 & 10 & 3 & 2 \\ \hline
$1024 \times 1024$ & 665 & 31 & 12 & 7 \\ \hline
$2048 \times 2048$ & $3.51 \times 10^3$ & 119 & 54 & 29 \\ \hline
\end{tabular}
\end{center}
\label{tbl:diff_time_once}
\end{table*}

\begin{table*}[htbp]
\caption{Calculation times of CGH calculation on the CPU and GPUs}
\begin{center}
\begin{tabular}{|c|c|c|c|c|}
\hline
 & CPU (ms) & \multicolumn{ 3}{c|}{GPU (ms)} \\ \hline
Number of PLSs & Intel Core i7 740QM & GeForce GTX 460M & GeForce GTX295 (1 chip) & GeForce GTX580 \\ \hline
248 & $1.3 \times 10^4$ & 87 & 67 & 31 \\ \hline
4596 & $2.2 \times 10^5$ & 650 & 562 & 230 \\ \hline
11646 & $5.9 \times 10^5$ & $1.7 \times 10^3$ & $1.43 \times 10^3$ & 579 \\ \hline
\end{tabular}
\end{center}
\label{tbl:cgh_time}
\end{table*}

\section{Applications to holography}
In this section. we show some applications to holography using the CWO++ library and its performances on the CPU and GPUs.

\subsection{Inline phase-only CGH (Kinoform)}

In this subsection, we show an example of generating an inline phase-only CGH, also known as kinoform.
A kinoform is calculated only by extracting the phase of a diffracted light onto the kinoform plane.
List \ref{lst:kinoform} shows the generation of an inline phase-only CGH with $512 \times 512$ pixels from the original image (Fig. \ref{fig:lena-diff} (a)).

In line 3, we add a random phase to the original image using the function ``SetRandPhase()" to spread the light.
The function automatically sets to the complex amplitude field (CWO\_FLD\_C\\
OMPLEX).
The random phase is generated by Xorshift RNGs algorithm \cite{rand}.
In lines 4 and 5, we calculate the kinoform by diffracting the original image at the propagation distance of 0.1 m, and subsequently extract the phase only from the diffracted light, which is the kinoform.

\begin{figure}[htb]
\centerline{
\includegraphics[width=9cm]{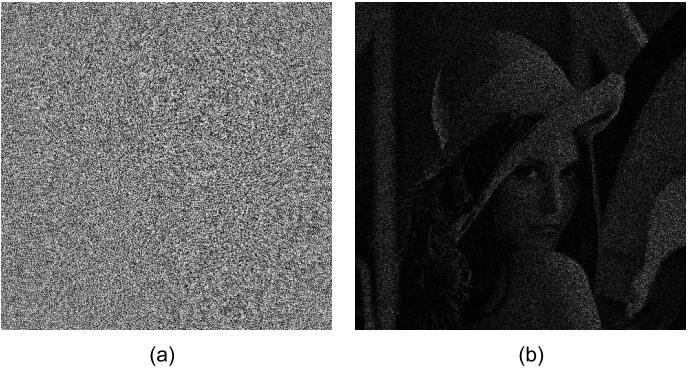}}
\caption{(a) Kinoform and (b) reconstructed image from the kinoform.}
\label{fig:kinoform}
\end{figure}

After line 7, these codes are for reconstruction from the kinoform.
In line 7, we convert the phase field (CWO\_FLD\_PHASE) to the complex amplitude field (CWO\_FLD\_COMPLEX).
In lines 8 to 11, we calculate the reconstructed image from the kinoform using the back propagation relative to the position of the original image.
Figure \ref{fig:kinoform} (a) and (b) show the kinoform pattern and the reconstructed image from the kinoform.

\begin{lstlisting}[caption={Inline phase-only CGH}, label=lst:kinoform, language=C, numbers=left, numberstyle=\tiny, stepnumber=1, frame=single]
CWO c;
c.Load("lena512x512.bmp");
c.SetRandPhase();
c.Diffract(0.1, CWO_FRESNEL_CONV);
c.Phase();

c.Complex();
c.Diffract(-0.1, CWO_FRESNEL_CONV);
c.Intensity();
c.Scale(255);
c.Save("kinoform_reconst.bmp");
\end{lstlisting}

\subsection{GS algorithm}

In the subsection, we implement the GS algorithm on the CPU and GPUs using the CWO++ library and show the performances of the algorithm on the CPU and GPUs.
Although it is possible to obtain a complete reconstructed image from a complex amplitude field, unfortunately, we lack an appropriate electric device to display the amplitude and phase of the complex amplitude field simultaneously.
Therefore, we need to select either the amplitude or the phase components of the complex amplitude field, which will cause the reconstructed image to deteriorate due to lack of information on the complex amplitude field. 
We employ the GS algorithm as an iterative algorithm \cite{GS, PingPong} in order to improve the deterioration of the reconstructed image.

\begin{figure*}[htb]
\centerline{
\includegraphics[width=14cm]{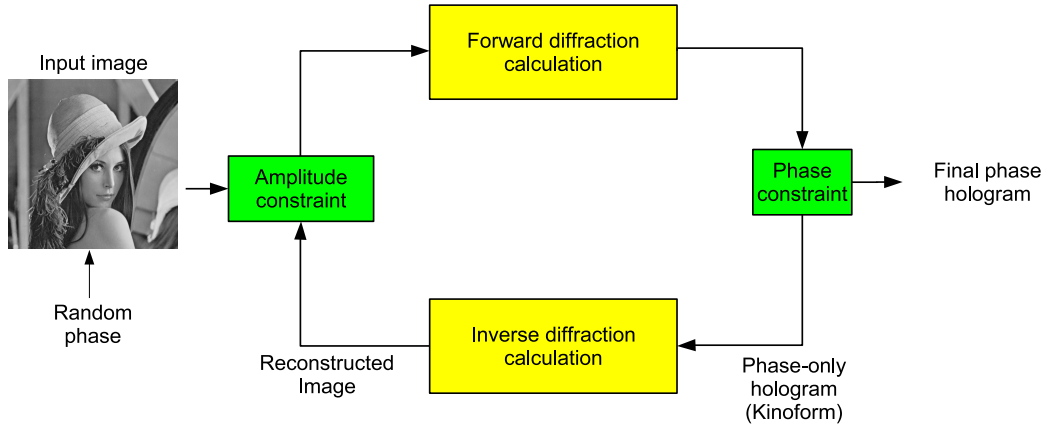}}
\caption{GS algorithm in Fresnel region in order to improve the deterioration of a reconstructed image.}
\label{fig:GS}
\end{figure*}

Figure \ref{fig:GS} shows typical a GS algorithm. 
In the GS algorithm for Fourier holograms, Fourier and inverse Fourier transforms correspond to reconstructions from a hologram and hologram generation, respectively. 
In the subsection, instead of Fourier and inverse Fourier transforms, we use the angular spectrum method and the back propagation of the same. 

We start the iteration by adding a random phase to an input image, and calculate the diffraction calculation from the latter.
We extract only the phase components (``Phase constraints" in Fig.\ref{fig:GS}) from the diffracted lights to generate a kinoform. 
The kinoforms are reconstructed by inverse diffraction calculation. 
We replace the amplitude of the reconstructed light with the original input image (``Amplitude constraint" in Fig.\ref{fig:GS}). 
Repeating the above processes, the GS algorithms gradually improve the quality of the reconstructed images.

List \ref{lst:GS_CPU} shows an example of the GS algorithm on a CPU.
In lines 1 to 7, we load an input image, calculate its square root, set a random phase to it, calculate the diffracted light to a kinoform plane, and subsequently generate a kinoform only by extracting the phase of the diffracted light.
The information of the original image is maintained in the instance ``a1", while instance ``a2" is used for the forward and back propagations.

In lines 8 to 17, we execute the iteration, the number of which is decided by ``ite\_num".
In lines 9 and 11, we calculate the reconstructed image by the back propagation of the angular spectrum method at a propagation distance of -0.1 m and extract only the phase information of the reconstructed light.
In line 13, we replace the amplitude of the reconstructed light with the original input image (``Amplitude constraint" in Fig.\ref{fig:GS}), where instances ``a1" and ``a2" hold the field type of CWO\_FLD\_INTE \\
NSITY and CWO\_FLD\_PHASE, respectively.

In lines 15 and 16, we recalculate a new kinoform from the new complex amplitude generated by line 13. Repeating the above processes, the GS algorithms gradually improve the quality of the reconstructed images.
In lines 19 to 23, we calculate a final reconstructed image from the kinoform.

\begin{lstlisting}[caption={GS algorithm on CPU.}, label=lst:GS_CPU, language=C, numbers=left, numberstyle=\tiny, stepnumber=1, frame=single]
CWO a1,a2;
a1.Load("lena2048x2048.bmp");
a1.Sqrt();
a2=a1;
a2.SetRandPhase();
a2.Diffract(0.1,CWO_ANGULAR);
a2.Phase();
for(int i=0;i<ite_num;i++){
	a2.Complex();
	a2.Diffract(-0.1,CWO_ANGULAR);
	a2.Phase();
	
	a2.Complex(a1,a2);

	a2.Diffract(0.1,CWO_ANGULAR);
	a2.Phase();
}
		
a2.Complex();
a2.Diffract(-0.1,CWO_ANGULAR);
a2.Intensity();
a2.Scale(255);
a2.Save("gs_on_cpu.bmp");
\end{lstlisting}

List \ref{lst:GS_GPU} shows an example of the GS algorithm on a GPU.
The example is almost the same as to List \ref{lst:GS_CPU}.
The iteration of lines 11 to 20 is executed on a GPU, so that the example will be calculated faster than the CPU version of List \ref{lst:GS_CPU}.

\begin{lstlisting}[caption={GS algorithm on GPU.}, label=lst:GS_GPU, language=C, numbers=left, numberstyle=\tiny, stepnumber=1, frame=single]
CWO c1,c2;
GWO g1,g2;
c1.Load("lena2048x2048.bmp");
c1.Sqrt();
c2=c1;
c2.SetRandPhase();
g1.Send(c1);
g2.Send(c2);
g2.Diffract0.1,CWO_ANGULAR);
g2.Phase();
for(int i=0;i<ite_num;i++){
	g2.Complex();
	g2.Diffract(-0.1,CWO_ANGULAR);
	g2.Phase();
	
	g2.Complex(g1,g2);

	g2.Diffract(0.1,CWO_ANGULAR);
	g2.Phase();
}
		
g2.Complex(); 
g2.Diffract(-0.1,CWO_ANGULAR);
g2.Intensity();
g2.Scale(255);
g2.Recv(c1);
c1.Save("gs_on_gpu.bmp");
\end{lstlisting}

Changing the resolution of the input image and the iteration number, we compare the calculation times of the GS algorithm on the CPU and GPUs, which are shown in Table \ref{tbl:gs}.
In the CPU calculations, we measured the time in lines 3 to 22 of List \ref{lst:GS_CPU}.
In the GPU calculations, we measured the time in lines 4 to 26 of List \ref{lst:GS_GPU}.
The calculation times on GPUs were much faster than those on the CPU.

\begin{figure*}[htb]
\centerline{
\includegraphics[width=14cm]{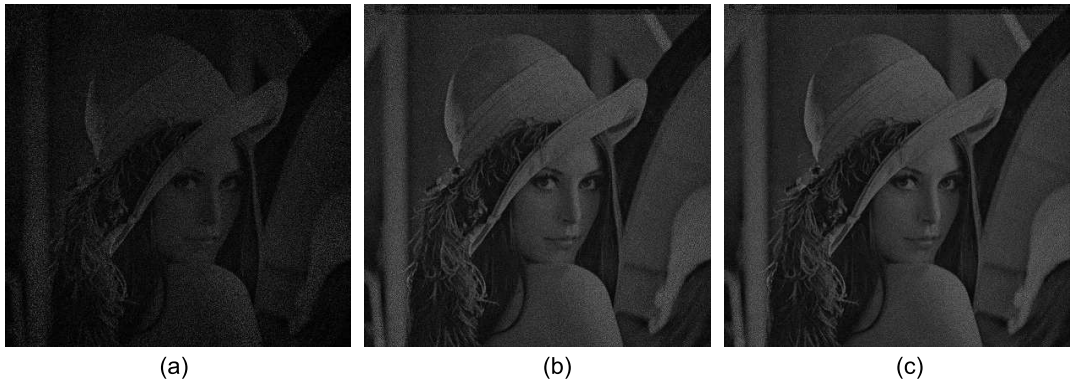}}
\caption{the reconstructed images when the resolution of the original image is $2,048 \times 2,048$ pixels and the numbers of iteration are 5, 20, and 40 respectively.}
\label{fig:GS_lena}
\end{figure*}

Figures \ref{fig:GS_lena} (a), (b) and (c) show the reconstructed images when the resolution of the original image was $2,048 \times 2,048$ pixels and the iteration numbers were 5, 20, and 40 respectively.

\begin{table*}[htbp]
\caption{Calculation times of the GS algorithm on the CPU and GPUs.}
\begin{center}
\begin{tabular}{|c|c|c|c|c|}
\hline
\multicolumn{ 1}{|c|}{Number of iterations} & CPU (ms) & \multicolumn{ 3}{c|}{GPU (ms)} \\ \cline{ 2- 5}
\multicolumn{ 1}{|c|}{} & Intel Core i7 740QM & GeForce GTX 460M & GeForce GTX295 (1 chip) & GeForce GTX580 \\ \hline
\multicolumn{ 5}{|c|}{Resolution of Input image : $512 \times 512$} \\ \hline
5 & $3.60 \times 10^3$ & 173 & 236 & 188 \\ \hline
10 & $6.53 \times 10^3$ & 285 & 273 & 210 \\ \hline
20 & $1.22 \times 10^4$ & 516 & 384 & 253 \\ \hline
40 & $2.36 \times 10^4$ & 984 & 482 & 338 \\ \hline
\multicolumn{ 5}{|c|}{Resolution of Input image : $1,024 \times1,024$} \\ \hline
5 & $1.59 \times 10^4$ & 539 & 920 & 731 \\ \hline
10 & $2.88 \times 10^4$ & 883 & $1.06 \times 10^3$ & 797 \\ \hline
20 & $5.56 \times 10^4$ & $1.59 \times 10^3$ & $1.36 \times 10^3$ & 930 \\ \hline
40 & $1.07 \times 10^5$ & $3.02 \times 10^3$ & $1.73 \times 10^3$ & $1.19 \times 10^3$ \\ \hline
\multicolumn{ 5}{|c|}{Resolution of Input image : $2,048 \times2,048$} \\ \hline
5 & $7.30 \times 10^4$ & $2.03 \times 10^3$ & $3.85 \times 10^3$ & $2.91 \times 10^3$ \\ \hline
10 & $1.33 \times 10^5$ & $3.36 \times 10^3$ & $4.25 \times 10^3$ & $3.16 \times 10^3$ \\ \hline
20 & $2.53 \times 10^5$ & $6.01 \times 10^3$ & $5.29 \times 10^3$ & $3.65 \times 10^3$ \\ \hline
40 & $4.98 \times 10^5$ & $1.12 \times 10^4$ & $7.17 \times 10^3$ & $4.65 \times 10^3$ \\ \hline
\end{tabular}
\end{center}
\label{tbl:gs}
\end{table*}

\section{Conclusion}
We developed the CWO++ library using the C++ class library to calculate the 2D and 3D diffraction and CGH calculations on CPU and GPU.
Our previous C-language based library, GWO, was not user-friendly because, for example, GWO library users have to manage the CPU and GPU memory allocation by themselves and so on.
The CWO++ library remains user-friendly by concealing troublesome programming within classes and the GPU calculation power while unaware of the GPGPU technique.
Applications capable of applying the CWO++ library cover a wide range of optics, ultrasonic and X-ray fields and so on.
In this paper, applications to holography are shown.
The CWO++ library will be distributed from {\color{blue} \bf \url{http://brains.te.chiba-u.jp/~shimo/cwo/}}.

\section*{Acknowledgments}
This research was partially supported by Japan Society for the Promotion of Science (JSPS), Grant-in-Aid for Young Scientists (B), 23700103, 2011, and, the Ministry of Internal Affairs and Communications, Strategic Information and Communications R\&D Promotion Programme (SCOPE)(09150542), 2009.

\appendix

\section{System requirements and installation}
\label{appendix_install}

System requirements for the CWO++ library are as follows:
\begin{enumerate}
\item OS : for Windows XP(32/ 64 bit), 7 
\item CUDA : CUDA 4.0 (32 bit version)  (if the GWO class or gwoPLS is used)
\end{enumerate}

The installation of the CWO++ library involves the following steps:
\begin{enumerate}
	\item Create a project file of Visual C++.
	\item Ensure the following dll and library files are placed in your project directory:
		\begin{enumerate}
			\item cwo.dll, cwo.lib, libfftw3f-3.dll (libfftw3f-3.dll can be download from Ref. \cite{fftw})
			\item gwo.dll, gwo.lib  (if you use GPU version of the CWO++ library)
		\end{enumerate}
\item Set library files (*.lib) to your VISUAL C++ project.
\item Set the following C++ and header files to your project:
	\begin{enumerate}
	\item cwo.h, cwo.cpp, cwo\_lib.h
	\item gwo.h, gwo.cpp, gwo\_lib.h  (if you use the GPU version of the CWO library)
	\end{enumerate}
\end{enumerate}

\section{Image formats}
\label{formats}
Using CImag library \cite{cimg}, CWO library allows us to read and write image data in the following formats:
\begin{enumerate}
\item Bitmap
\item Jpeg
\item Png
\item Tiff
\end{enumerate}
If you read and write image formats other than bitmap format, you must install ImageMagik \cite{imagemagik} on your computer.

\begin{thebibliography}{00}

\bibitem{goodman}
J.W.Goodman, ``Introduction to Fourier Optics (3rd ed.),'' Robert \& Company (2005). 
\bibitem{okan} 
Okan K. Ersoy, ``Diffraction, Fourier Optics And Imaging,'' Wiley-Interscience (2006).
\bibitem{ultrasonic}
E. G. Williams, ``Fourier Acoustics -- Sound Radiation and Nearfield Acoustical Holography,'' Academic Press (1999).
\bibitem{xray}
D.M. Paganin, ``Coherent X-Ray Optics,'' Oxford University Press (2006).
\bibitem{slinger}
C. Slinger, C. Cameron, M. Stanley, M, ``Computer-Generated Holography as a Generic Display Technology,'' Computer {\bf 38}, 46--53 (2005).
\bibitem{bentonbook}
S. A. Benton et al., ``Holographic Imaging,'' Wiley-Interscience (2008).

\bibitem{nlut}
S.C. Kim and E.S. Kim, ``Effective generation of digital holograms of three-dimensional objects using a novel look-up table method,''  Appl. Opt. {\bf 47},  D55--D62 (2008) .

\bibitem{sakamoto}
H. Sakata and Y. Sakamoto, ``Fast computation method for a Fresnel hologram using three-dimensional affine transformations in real space,'' Appl. Opt. {\bf 48}, H212--H221 (2009).

\bibitem{image_hol} 
H. Yoshikawa, T. Yamaguchi, and R. Kitayama, ``Real-Time Generation of Full color Image Hologram with Compact Distance Look-up Table,''  OSA Topical Meeting on Digital Holography and Three-Dimensional Imaging 2009, DWC4 (2009). 

\bibitem{ExtreamCGH}
K. Matsushima and S. Nakahara, ``Extremely high-definition full-parallax computer-generated hologram created by the polygon-based method,'' Appl. Opt. {\bf 48}, H54--H63 (2009).

\bibitem{FraunhoferCGH}
Y. Liu, J. Dong, Y. Pu, H. He, B. Chen, H. Wang, H. Zheng and Y. Yu, ``Fraunhofer computer-generated hologram for diffused 3D scene in Fresnel region,'' Opt. Lett. {\bf 36}, 2128--2130 (2011). 

\bibitem{schnars1}
U.Schnars and W. Juptner, ``Direct recording of holograms by a CCD target and numerical Reconstruction,'' Appl.Opt., {\bf 33,} 2, 179--181 (1994).
\bibitem{schnars2}
U.Schnars and W.Jueptner, ``Digital Holography - Digital Hologram Recording, Numerical Reconstruction, and Related Techniques,'' Springer (2005).
\bibitem{kim}
M. K. Kim, ``Principles and techniques of digital holographic microscopy,'' SPIE Reviews {\bf 1}, 018005 (2010).
\bibitem{DHM2}
M. Gustafsson, M. Sebesta, B. Bengtsson, S. G. Pettersson, P. Egelberg, and T. Lenart, ``High-resolution digital transmission microscopy: a Fourier holography approach,'' Opt. Lasers Eng. {\bf 41}, 553--563 (2004).


\bibitem{masuda1}
N. Masuda, T. Ito, K. Kayama, H. Kono, S. Satake, T. Kunugi and Kazuho Sato, ``Special purpose computer for digital holographic particle tracking velocimetry,'' Opt. Express {\bf 14}, 587--592 (2006).
\bibitem{satake}
S. Satake, H. Kanamori, T. Kunugi, K. Sato, T. Ito, and K. Yamamoto, ``Parallel computing of a digital hologram and particle searching for microdigital-holographic particle-tracking velocimetry,'' Appl. Opt. {\bf 46}, 538--543 (2007).

\bibitem{GS} 
J. R. Fienup, `` Phase retrieval algorithms: a comparison, '' Appl. Opt. {\bf 21}, 2758--2769 (1982).

\bibitem{PingPong} 
R. G. Dorsch, A. W. Lohmann, and S. Sinzinger, ``Fresnel ping-pong algorithm for two-plane computer-generated hologram display, '' Appl. Opt. {\bf 33}, 869--875 (1994). 

\bibitem{YG}
G. Yang, B. Dong, B. Gu, J. Zhuang, and O. K. Ersoy, ``Gerchberg-Saxton and Yang-Gu algorithms for phase retrieval in a nonunitary transform system: a comparison, '' Appl. Opt. {\bf 33}, 209--218 (1994).

\bibitem{wave_reconst1}
G. Pedrini, W. Osten, and Y. Zhang, ``Wave-front reconstruction from a sequence of interferograms recorded at different planes, '' Opt. Lett. {\bf 30}, 833--835 (2005).

\bibitem{wave_reconst2}
D. Zheng, Y. Zhang, J. Shen, C. Zhang and G. Pedrini, `` Wave field reconstruction from a hologram sequence, ''Opt. Communications {\bf 249} 73--77 (2005). 

\bibitem{wave_reconst3}
A. Grjasnow, A. Wuttig and R. Riesenberg, `` Phase resolving microscopy by multi-plane diffraction detection, '' J. Microscopy {\bf 231}, 115--123 (2008).

\bibitem{proj1} 
E. Buckley, ``Holographic Laser Projection, ''  J. Display Technol. {\bf 99}, 1--6 (2010).
\bibitem{proj2} 
E. Buckley, ``Holographic projector using one lens, '' Opt. Lett. {\bf 35}, 3399--3401 (2010). 
\bibitem{proj3}
M. Makowski, M. Sypek, and A. Kolodziejczyk, ``Colorful reconstructions from a thin multi-plane phase hologram, ''   Opt. Express {\bf 16}, 11618--11623 (2008).
\bibitem{proj4} 
M. Makowski, M. Sypek, I. Ducin, A. Fajst, A. Siemion, J. Suszek, and A. Kolodziejczyk, ``Experimental evaluation of a full-color compact lensless holographic display, '' Opt. Express {\bf 17}, 20840--20846 (2009). 
\bibitem{shimo_proj1}
T. Shimobaba, T. Takahashi, N. Masuda, and T. Ito, ``Numerical study of color holographic projection using space-division method,'' Opt. Express {\bf 19}, 10287-10292 (2011) .
\bibitem{shimo_proj2}
T. Shimobaba, A. Gotchev, N. Masuda and T. Ito, ``Proposal of zoomable holographic projection method without zoom lens,'' IDW'11 (The 18th international Display Workshop) (to be appeared in Dec. 2011).


\bibitem{encryption1}
O. Matoba and B. Javidi, ``Encrypted optical memory system using three-dimensional keys in the Fresnel domain, '' Opt. Lett. {\bf 24}, 762--764 (1999).
\bibitem{encryption2}
E. Tajahuerce and B. Javidi, ``Encrypting three-dimensional information with digital holography, '' Appl. Opt. {\bf 39}, 6595--6601 (2000).
\bibitem{encryption3}
B. Javidi and Takanori Nomura, ``Securing information by use of digital holography, '' Opt. Lett. {\bf 25}, 28--30 (2000). 
\bibitem{steganography}
H. Hamam, ``Digital holography-based steganography, '' Opt. Lett. {\bf 35}, 4175--4177 (2010).

\bibitem{3d} 
R. Piestunand and J. Shamir, ``Synthesis of three-dimensional light fields and applications,'' Proc. IEEE {\bf 90}, 222--244 (2002). 
\bibitem{mems}
T. P. Kurzweg, S. P. Levitan, P. J. Marchand, J. A. Martinez, K. R. Prough, D. M. Chiarulli, ``A CAD Tool for Optical MEMS, '' Proc.36th ACM/IEEE conf. on Design automation, 879--884 (1999).
\bibitem{mems2}
T. P. Kurzweg, S. P. Levitan, J. A. Martinez, M. Kahrs, D. M. Chiarulli, ``An Efficient Optical Propagation Technique for Optical MEM Simulation, '' Fifth International Conference on Modeling and Simulation of Microsystems (MSM2002), 352--355 (2002).



\bibitem{horn1}
T. Ito, T. Yabe, M. Okazaki and M. Yanagi, ``Special-purpose computer HORN-1 for reconstruction of virtual image in three dimensions,'' Comput.Phys.Commun. {\bf 82}, 104--110 (1994).
\bibitem{horn2}
T. Ito, H. Eldeib, K. Yoshida, S. Takahashi, T. Yabe and T. Kunugi, ``Special-Purpose Computer for Holography HORN-2,'' Comput.Phys.Commun. {\bf 93}, 13--20 (1996).
\bibitem{horn3}
T.Shimobaba, N.Masuda, T.Sugie, S.Hosono, S.Tsukui and T.Ito, ``Special-Purpose Computer for Holography HORN-3 with PLD technology,'' Comput. Phys. commun. {\bf 130}, pp. 75--82, (2000).
\bibitem{horn4}
T. Shimobaba, S. Hishinuma and T.Ito, ``Special-Purpose Computer for Holography HORN-4 with recurrence algorithm,'' Comput. Phys. Commun. {\bf 148}, 160--170 (2002).
\bibitem{horn5}
T. Ito, N. Masuda, K. Yoshimura, A. Shiraki, T. Shimobaba and T. Sugie, ``A special-purpose computer HORN-5 for a real-time electroholography,'' Opt. Express {\bf 13} 1923-1932 (2005).
\bibitem{horn6}
Y. Ichihashi, H. Nakayama, T. Ito, N. Masuda, T. Shimobaba, A. Shiraki and T. Sugie, ``HORN-6 special-purpose clustered computing system for electroholography,'' Opt. Express {\bf 17}, 13895--13903 (2009). 

\bibitem{masuda2}
Y. Abe, N. Masuda, H. Wakabayashi, Y. Kazo, T. Ito, S. Satake, T. Kunugi and K. Sato, ``Special purpose computer system for flow visualization using holography technology,'' Opt. Express {\bf 16}, 7686--7692 (2008).

\bibitem{gwo}
T. Shimobaba, T. Ito, N. Masuda, Y. Abe, Y. Ichihashi, H. Nakayama, N. Takada, A.Shiraki and T. Sugie, ``Numerical calculation library for diffraction integrals using the graphic processing unit: the GPU-based wave optics library,'' Journal of Optics A: Pure and Applied Optics, 10, 075308, 5pp, (2008).


\bibitem{realtimeDHM}
T.Shimobaba, Y.Sato, J.Miura, M.Takenouchi, and T.Ito, ``Real-time digital holographic microscopy using the graphic processing unit,'' Opt. Express 16, 11776-11781 (2008)
\bibitem{cad}
T.Shimobaba, J.Miura and T.Ito, ``A computer aided design tool for developing an electroholographic display,'' Journal of Optics A: Pure and Applied Optics, 11, 085408 (5pp) (2009)
\bibitem{wrp1}
T. Shimobaba, N. Masuda and T. Ito, ``Simple and fast calclulation algorithm for computer-generated hologram with wavefront recording plane,'' Opt. Lett. {\bf 34}, 3133--3135 (2009).
\bibitem{wrp2}
T. Shimobaba, H. Nakayama, N. Masuda and T. Ito, ``Rapid calculation of Fresnel computer-generated-hologram using look-up table and wavefront-recording plane methods for three-dimensional display,'' Optics Express, 18, 19, 19504-19509 (2010).
\bibitem{multiDHM}
T. Shimobaba, N. Masuda, Y. Ichihashi and T. Ito, ``Real-time digital holographic microscopy observable in multi-view and multi-resolution,'' Journal of Optics, 12, 065402 (4pp) (2010)
\bibitem{airy1}
H. T. Dai, X. W. Sun, D. Luo, and Y. J. Liu, ``Airy beams generated by a binary phase element made of polymer-dispersed liquid crystals,'' Opt. Express 17, 19365-19370 (2009) 
\bibitem{airy2}
D. Luoa, H.T . Dai , X. W. Sun, and H. V. Demira, ``Electrically switchable finite energy Airy beams generated by a liquid crystal cell with patterned electrode,'' Optics Communications, 283, 3846-3849 (2010).


\bibitem{shift_fre1}
R. P. Muffoletto, J. M. Tyler, and J. E. Tohline, ``Shifted Fresnel diffraction for computational holography,'' Opt. Express {\bf 15}, 5631--5640 (2007).\bibitem{scaled_fourier}
D. H. Bailey and P. N. Swarztrauber, ``The Fractional Fourier Transform and Applications,'' SIAM Review {\bf 33}, 389--404 (1991).

\bibitem{shift_fre2}
M. Leutenegger, R. Rao, R. A. Leitgeb, and T. Lasser, ``Fast focus field calculations,'' Opt. Express {\bf 14}, 11277--11291 (2006).
\bibitem{shift_fre3}
J. F. Restrepo and J. Garcia-Sucerquia, ``Magnified reconstruction of digitally recorded holograms by Fresnel Bluestein transform,''  Appl. Opt. {\bf 49}, 6430--6435 (2010). 
\bibitem{shift_angular}
K. Matsushima, ``Shifted angular spectrum method for off-axis numerical propagation,'' Opt. Express {\bf 18}, 18453--18463 (2010). 
\bibitem{band_angular}
K. Matsushima and T. Shimobaba, ``Band-limited angular spectrum method for numerical simulation of free-space propagation in far and near fields,'' Opt. Express {\bf 17}, 19662--19673 (2009).
\bibitem{fftw} 
FFTW Home Page, \url{http://www.fftw.org/}.

\bibitem{3d_direct}
J. Lin, X.-C. Yuan, S. S. Kou, C. J. R. Sheppard, O. G. Rodriguez-Herrera and J. C. Dainty, ``Direct calculation of a three-dimensional diffracted field,'' Opt. Lett. {\bf 36}, 1341--1343 (2011).

\bibitem{Ahrenberg}
L. Ahrenberg, P. Benzie, M. Magnor and J. Watson, ``Computer generated holograms from three dimensional meshes using an analytic light transport model,''  Appl. Opt. {\bf 47}, 1567--1574 (2008).

\bibitem{lut} 
M. Lucente, ``Interactive Computation of holograms using a Look-up Table,''  J. Electron. Imaging {\bf 2}, 28--34 (1993).

\bibitem{lucenteGPU}
M. Lucente and T. A. Galyean, ``Rendering Interactive Holographic Images,'' Proc. of SIGGRAPH 95  387--394 (1995).  
\bibitem{masuda3}
N. Masuda, T. Ito, T. Tanaka, A. Shiraki and T. Sugie, ``Computer generated holography using a graphics processing unit,''  Opt. Express {\bf 14,} 587--592 (2006).
\bibitem{Ahrenberg}
L. Ahrenberg, P. Benzie, M. Magnor, J. Watson, ``Computer generated holography using parallel commodity graphics hardware,'' Opt. Express {\bf 14,} 7636--7641 (2006).
\bibitem{Onural}
H. Kang, F. Yaras, and L. Onural, ``Graphics processing unit accelerated computation of digital holograms,'' Appl. Opt. {\bf 48}, H137--H143 (2009).
\bibitem{lutgpu} 
Y. Pan, X. Xu, S. Solanki, X. Liang, R. Bin A. Tanjung, C. Tan, and T. C. Chong, ``Fast CGH computation using S-LUT on GPU,''  Opt. Express {\bf 17}, 18543--18555 (2009).

\bibitem{DHAhrenberg} 
L. Ahrenberg, A. J. Page, B. M. Hennelly, J. B. McDonald, and T. J. Naughton, ``Using Commodity Graphics Hardware for Real-Time Digital Hologram View-Reconstruction,'' J. Display Technol. {\bf 5}, 111--119 (2009).

\bibitem{DHCarl} 
D. Carl, M. Fratz, M. Pfeifer, D. M. Giel, and H. Hofler, ``Multiwavelength digital holography with autocalibration of phase shifts and artificial wavelengths,'' Appl. Opt. {\bf 48}, H1--H8 (2009).

\bibitem{DHNaughton} 
N. Pandey, D. P. Kelly, T. J. Naughton and B. M. Hennellyr, ``Speed up of Fresnel transforms for digital holography using pre-computed chirp and GPU processing,'' Proc. SPIE {\bf 7442}, 744205 (2009). 

\bibitem{DHGarcia} 
C. Trujillo, John F. Restrepo and J. Garcia-Sucerquia, ``Real time numerical reconstruction of digitally recorded holograms in digital in-line holographic microscopy by using a graphics processing unit,'' Photonics Letters of Poland {\bf 2}, 177--179 (2010).

\bibitem{amd_cgh}
T. Shimobaba, T. Ito, N. Masuda, Y. Ichihashi, and N. Takada, ``Fast calculation of computer-generated-hologram on AMD HD5000 series GPU and OpenCL,'' Opt. Express {\bf 18}, 9955--9960 (2010).
\bibitem{amd_diff}
T. Nishitsuji, T. Shimobaba, T. Sakurai, N. Takada, N. Masuda, and T. Ito, ``Fast calculation of Fresnel diffraction calculation using AMD GPU and OpenCL,'' in Digital Holography and Three-Dimensional Imaging, OSA Techinal Digest (CD) (Optical Society of America, 2011), paper DWC20. 

\bibitem{rand}
G. Marsaglia, ``Xorshift RNGs,'' Journal of Statistical Software {\bf 8}, 1--6 (2003).


\bibitem{cimg}
CImg homepage, \url{http://cimg.sourceforge.net/}

\bibitem{imagemagik}
ImageMagik homepage, \url{http://www.imagemagick.org/script/index.php}

\end{thebibliography}
\end{document}
\endinput